\documentclass{emulateapj}
\shorttitle{X-ray properties of maxBCG clusters}
\shortauthors{Rykoff et al.}

\newcommand{\kpc}{h^{-1}\,\mathrm{kpc}}
\newcommand{\Mpc}{h^{-1}\,\mathrm{Mpc}}
\newcommand{\lum}{h^{-2}\,\mathrm{ergs}\,\mathrm{s}^{-1}}
\newcommand{\lmean}{\bar{L}_X}
\newcommand{\nmean}{\bar{N}_{200}}
\newcommand{\lmed}{\widetilde{L}_X}
\newcommand{\zmed}{\tilde{z}}
\newcommand{\lnlmean}{\overline{\ln{L_X}}}
\newcommand{\tmean}{\bar{T}_X}
\newcommand{\siglnl}{\sigma_{\ln{L}}}
\newcommand{\Bgc}{B_\mathrm{gc}}
\newcommand{\smed}{\widetilde{\sigma}}
\newcommand{\smedfull}{\exp(\overline{\ln{\sigma}})}
\newcommand{\lbcg}{L_{\mathrm{BCG}}}
\newcommand{\lbcgmean}{\bar{L}_{\mathrm{BCG}}}
\newcommand{\lxbolo}{L_{X,\mathrm{bolo}}}
\newcommand{\up}{$u$}
\newcommand{\gp}{$g$}
\newcommand{\rp}{$r$}
\newcommand{\ip}{$i$}
\newcommand{\zp}{$z$}

\begin{document}

\title{Measuring the mean and scatter of the X-ray luminosity -- optical 
richness relation for maxBCG galaxy clusters}

\author{
E.~S.~Rykoff\altaffilmark{1,2},
T.~A.~McKay\altaffilmark{2,3,4},
M.~R.~Becker\altaffilmark{2},
A.~Evrard\altaffilmark{2,3,4},
D.~E.~Johnston\altaffilmark{5},
B.~P.~Koester\altaffilmark{6},
E.~Rozo\altaffilmark{7},
E.~S.~Sheldon\altaffilmark{8},
R.~H.~Wechsler\altaffilmark{9}
}
\altaffiltext{1}{TABASGO Fellow, Physics Department, University of California
  at Santa Barbara, 3019 Broida Hall, Santa Barbara, CA 93106, erykoff@physics.ucsb.edu}
\altaffiltext{2}{Physics Department, University of Michigan, Ann Arbor, MI 48109}
\altaffiltext{3}{Astronomy Department, University of Michigan, Ann Arbor, MI 48109}
\altaffiltext{4}{Michigan Center for Theoretical Physics, Ann Arbor, MI 48109}
\altaffiltext{5}{Jet Propulsion Laboratory, 4800 Oak Grove Drive, Pasadena, CA
        91109}
\altaffiltext{6}{Department of Astronomy, The University of Chicago, Chicago,
        IL 60637}
\altaffiltext{7}{CCAPP Fellow, The Ohio State University, Columbus, OH 43201}
\altaffiltext{8}{Center for Cosmology and Particle Physics, Physics Department, New York University, New York, NY 10003}
\altaffiltext{9}{Kavli Institute for Particle Astrophysics and Cosmology,
        Physics Department and Stanford Linear Accelerator Center, Stanford
        University, 382 Pueblo Mall, Stanford, CA 94305}

\begin{abstract}

Determining the scaling relations between galaxy cluster observables requires
large samples of uniformly observed clusters.  We measure the mean X-ray
luminosity--optical richness ($\lmean$--$\nmean$) relation for an approximately
volume-limited sample of more than 17,000 optically-selected clusters from the
maxBCG catalog spanning the redshift range $0.1<z<0.3$.  By stacking the X-ray
emission from many clusters using \emph{ROSAT} All-Sky Survey data, we are able
to measure mean X-ray luminosities to $\sim$10\% (including systematic errors)
for clusters in nine independent optical richness bins. In addition, we are
able to crudely measure individual X-ray emission from $\sim 800$ of the
richest clusters.  Assuming a log-normal form for the scatter in the
$L_X$--$N_{200}$ relation, we measure $\sigma_{\ln{L}}=0.86\pm0.03$ at fixed
$N_{200}$.  This scatter is large enough to significantly bias the mean stacked
relation.  The corrected median relation can be parameterized by $\lmed =
e^{\alpha}(\nmean/40)^{\beta}\,10^{42}\,\lum$, where $\alpha = 3.57 \pm 0.08$
and $\beta = 1.82\pm0.05$.  We find that X-ray selected clusters are
significantly brighter than optically-selected clusters at a given optical
richness.  This selection bias explains the apparently X-ray underluminous
nature of optically-selected cluster catalogs.
\end{abstract}

\keywords{galaxies: clusters -- X-rays: galaxies: clusters}

\section{Introduction}

Galaxy clusters, as the largest peaks in the cosmic density field, play an
important role in astrophysics and cosmology \citep[e.g.][]{b03,v05}. 
Structure formation theory, realized in large scale N-body simulations, 
makes robust predictions for cluster space density and clustering within
various cosmological models. Since clusters are also the most observationally
accessible features of large scale structure, they provide an opportunity 
to place
strong constraints on both cosmological parameters and the growth of structure.
The great challenge of cluster cosmology lies in confidently relating the
dark matter halos we can robustly predict to the baryonic structures we 
observe. Substantial work is being done to close the gap between theory and 
observations from both sides.

On the theory side, numerical simulations of ever increasing 
complexity and resolution provide new insights into the
evolution of baryons within clusters, and to elucidate the connection
between cluster galaxies and dark matter substructure \citep{nk05, cwk06}. 
Observationally, much work is being done to assemble large samples 
of clusters detected and observed in a wide variety of ways. This is possible
because galaxy clusters provide a rich suite of observables. Optical light 
is emitted by individual cluster galaxies as well as intracluster stars. 
X-rays are emitted by both the hot intracluster medium (ICM) and AGN within 
cluster galaxies. This same hot ICM scatters microwave
background photons passing through the cluster, distorting their spectrum.
Finally, the total projected mass distribution of the cluster produces 
weak and sometimes strong lensing distortions in the images of background
galaxies. Each of these observables presents an opportunity to detect 
clusters and measure their properties. When combined, they allow us to 
cross-check our understanding of cluster physics in a variety of ways.

Clusters were first detected as early as the 18th century as anomalous 
groupings of similarly bright galaxies \citep{biv00}. Optical surveys, which 
are relatively inexpensive, have long provided the largest cluster catalogs, 
primarily because of their ability to detect objects with relatively low 
mass thresholds. While successful in identifying thousands of clusters 
\citep{abe58, aco89} and providing the first evidence for dark matter 
\citep{zwi33, zwi37}, early optical detection was plagued by projection 
of galaxies along the line of sight \citep{cgnl95}. Precise CCD photometry 
has enabled searches for galaxies clustered in space, brightness, and 
color~\citep[for a review of optical selection, see][]{gal06}, 
substantially reducing the problems of projection. These more recent 
optical surveys also naturally provide accurate photometric redshifts.

With the advent of X-ray satellites, detection of thermal emission from the hot
ICM became possible \citep{gkmlt71}.  Because X-ray emission depends on the
square of the density, this provides a higher contrast that is largely immune
to projection effects. But when spatial resolution is low, contamination from
non-thermal X-ray sources can be a difficult problem. These can be point 
sources like AGN, the non-thermal emission from a cooling core, or ongoing 
merger activity that has thrown a cluster out of thermal equilibrium.  Over 
the last several decades, X-ray surveys have been used to assemble a large 
number of cluster catalogs \citep{gl94, ebveh96, rdng98, rnhup00, bvhmg00, 
mmqvh03, bsgcv04, bvheq06}.  Much of our understanding of cluster physics 
today derives from these X-ray selected catalogs.

Substantial effort has gone into comparing optically and X-ray selected
catalogs \citep{bah77, es91, dmslp01, dsmlp02, mdmb03, ye03, gbcz04, pbbvy04},
and to comparing optical and X-ray properties of clusters to their weak lensing
and SZ signals \citep{skbfw96, all98, fis99, zsdeh01, dkilm02, cskc04, dsbl05,
mfbgh05, lbcjn06, hoe07, mhbsm07, bskce07}. Comprehensive comparisons of 
optical and X-ray properties of clusters have been hampered by the lack of 
large samples that are uniformly observed in both passbands.

In this paper we describe measurements of the X-ray properties of the largest
publicly available optically-selected cluster sample: the maxBCG catalog
\citep{kmawe07b}. This approximately volume-limited cluster catalog spans the
redshift range from $0.1<z<0.3$. Uniform optical photometry and relatively
precise photometric redshifts ($\Delta_{z} \le 0.015$) for all these clusters
are available from the same Sloan Digital Sky Survey~\citep[SDSS: ][]{yaaaa00}
data from which the clusters were selected. X-ray observations of all these
maxBCG clusters are available from the \emph{ROSAT} All-Sky Survey~\citep[RASS:
][]{vabbb99}.  While RASS exposures are too shallow to allow significant
individual detections of every maxBCG cluster, they provide precise
measurements of the mean X-ray luminosity ($\lmean$) as a function of optical
richness and redshift.  In addition, the low signal-to-noise measurements of
X-ray emission from individual clusters can be used both to confirm the
measurement of mean X-ray emission obtained by stacking and to provide
estimates of the scatter in the optical richness--X-ray luminosity relation.

The maxBCG catalog has been studied in a variety of complementary ways. For
example, both dynamical~\citep[][henceforth B07]{bmkwr07} and weak
lensing~\citep[S07, J07]{she07,joh07} observations of these clusters have 
been extracted
from SDSS data. These earlier observations can be combined with the mean
$\lmean$ measurements presented here to provide further insight into cluster
physics.  The $\lmean$--$\sigma$ relation inferred for maxBCG clusters is
described here, while the corresponding $\lmean$--$M_{200}$ relation obtained
from weak lensing is discussed in a companion letter~\citep{ryk07}.

An analysis similar to that reported in this paper was performed for the NIR
selected 2MASS cluster catalog by \citet*[][henceforth DKM07]{dkm07}.  The
2MASS NIR flux-limited catalog has $\sim4000$ nearby ($\bar{z}\sim 0.02$)
groups and clusters ranging in mass from $\sim10^{13}-10^{15}\,M_\sun$,
selected with a matched-filter algorithm~\citep{kwhmj03}.  DKM07 are the first
to take a large optically-selected cluster catalog and measure the mean
(stacked) X-ray properties using RASS, rather than simply cross-correlating
optically-selected and X-ray-selected clusters.  They find that the X-ray
luminosity of 2MASS clusters scales with optical richness (their $N_{*666}$,
the number of galaxies brighter than $L_*$ within $\sim R_{200}$).  In
addition, they derive X-ray temperatures and hydrostatic masses for the stacked
2MASS clusters.

In Section \ref{sec: input data}, we briefly review the SDSS data, maxBCG
catalog, and RASS data on which this study is based. Section \ref{sec: x-ray
analysis} describes our measurement of X-ray luminosities for individual
clusters, as well as our methods for determining the mean X-ray luminosity of a
set of clusters with similar richness. We describe the mean relation of X-ray
luminosity as a function of richness, scatter in this relation, and the 
underlying median $\lmed - \nmean$ relation in Section \ref{sec:meanrelation}. 
Section \ref{sec:biases} discusses several possible sources of systematic 
bias in these results. In Section \ref{sec:lsig} we combine these results with
dynamical measurements of maxBCG clusters to produce a measurement of 
the $\lmed - \smed$ relation. Conclusions and some discussion of
future work are presented in Section \ref{sec:summary}. Throughout this work
we use a $\Lambda$CDM cosmology with $H_0 =
100\,h\,\mathrm{km}\,\mathrm{s}^{-1}$ and $\Omega_m = 1 - \Omega_\Lambda =
0.3$.

\section{Input data} \label{sec: input data}

The measurements described here are based on two wide-area imaging surveys;
SDSS and RASS.  Galaxy clusters are selected from the SDSS
five-band imaging data using a red sequence selection method. SDSS data also
allow measurement of cluster redshifts and richnesses.  X-ray emission from
these clusters is then measured by from the RASS photon maps. In this section
we describe briefly the SDSS imaging data, galaxy cluster selection and
calibration, and RASS input data.

\subsection{SDSS data}

Optical data for this study are drawn from Sloan Digital Sky 
Survey\footnote{http://www.sdss.org}: a combined imaging and spectroscopic survey 
of 10$^4$ deg$^2$ in the North Galactic Cap and a smaller region in the 
South. The imaging survey was carried out in drift-scan mode in five SDSS 
filters (\up, \gp, \rp, \ip, \zp) to a limiting magnitude of $r<22.5$ 
\citep{figds96, gcrsb98, stkrf02}. Photometric errors are typically limited 
at bright magnitudes by systematic uncertainties at the 3\% level. The 
spectroscopic survey targets both a ``main'' sample of galaxies with 
$r<17.8$ and a median redshift of $z\sim0.1$ \citep{swlna02} and 
a ``luminous red galaxy'' sample \citep{eagsc01} which is approximately 
volume limited out to z=0.38. For more details of the SDSS see 
\citet{yaaaa00} and \citet{a07}.

\subsection{\label{sec:maxbcg}maxBCG Catalog}

The maxBCG cluster catalog is selected from imaging data contained in DR5 of
the SDSS. Selection of galaxy clusters from this imaging data is done using the
``maxBCG'' algorithm. Details of the algorithm are presented in
\citet{kmawe07a}, while the catalog and a description of its properties may be
found in \citet{kmawe07b}.  In brief, the algorithm exploits two well-known
features of rich galaxy clusters in addition to the tight spatial clustering of
cluster galaxies.  First, the bright end of the cluster luminosity function is
dominated by red sequence galaxies with a small dispersion in color-magnitude
space (the E/S0 ridgeline).  Second, clusters contain a distinct brightest
cluster galaxy (BCG) located near the center of the galaxy distribution. While
some clusters lack an obvious central, dominant galaxy, every cluster does
possess some red sequence galaxy brighter than any other.

The algorithm measures independently the likelihood that a galaxy is spatially
located in an overdensity of E/S0 ridgeline galaxies with similar $g-r$ and
$r-i$ colors, and that it has the color and magnitude properties of a typical
BCG. Both likelihoods are evaluated for every SDSS
galaxy at a grid of redshifts. The redshift which maximizes the product of
these likelihoods is then found for each galaxy. For E/S0 galaxies, this
corresponding maximum likelihood redshift then provides a good estimate of the
cluster redshift.

Once this list of cluster center likelihoods is assembled, these potential
centers are ranked by decreasing maximum likelihood. The first cluster is
seeded on the highest likelihood center, that cluster's BCG. Galaxies projected
within a scaled radius, $R_{200}$, of this BCG and within $\pm 0.02$ in $z$ are
eliminated from the list of potential centers.  $R_{200}$ is the radius
interior to which the mean density is 200 times the critical density
($\rho_{crit}$) as determined from SDSS galaxy populations~\citep{hmwas05}.
The process is repeated for the next most likely BCG on the list, given that it
has not been eliminated by a higher likelihood BCG, and continues likewise
until all potential centers have either been labeled as cluster BCGs or have
been subsumed by higher likelihood centers. Each cluster defined in this way
has a center defined as the BCG location, an estimated redshift, and a
richness, $N_{200}$, given by the number of E/S0 ridgeline members falling
within $R_{200}$ of the BCG and brighter than 0.4 $L_{*}$. The final cluster
catalog contains an array of measured properties, including photometric
redshifts, richnesses, optical luminosities, and locations.

The public maxBCG catalog contains a total of 13,823 clusters drawn from
approximately 7500 square degrees of sky between redshifts of 0.1 and 0.3, with
a median redshift $\zmed=0.23$. The center for each cluster is defined as the
location of the BCG identified by the algorithm. The richness of the cluster,
$N_{200}$, ranges between 10 and 188 in the public catalog, and in principle
extends down to $N_{200}=1$ where the maxBCG selection function is less
well-understood. For this study we include some clusters of lower richness,
adding an additional 3532 clusters with $N_{200} = 9$.  This slightly extended
catalog allows us to use the same richness bins studied in the analysis of
maxBCG galaxy dynamics~(B07) and gravitational lensing~(S07, J07).

Redshift estimates for the clusters, produced as part of the cluster finding
process, have been shown by comparison to spectroscopic redshifts to be quite
accurate, with $\Delta_{z} \le 0.015$ \citep{kmawe07b}. Catalog completeness
and purity have been studied in some detail in \citet{rwkme07a}; both are quite
high.  Completeness is estimated to be $\ge 90\%$ for clusters with masses
greater than $10^{14}\,h^{-1}\,M_{\sun}$, and purity is $\ge 90\%$ for clusters
with richnesses $N_{200} \ge 10$.

The cluster population in this catalog has been used to derive constraints on
cosmology \citep{rwkme07a} using cluster counts. The relationship between the
maxBCG richness $N_{200}$ and mass has been studied through galaxy
dynamics~(B07) and weak lensing (S07, J07). Further work on galaxy
populations, mass-to-light ratios, and improved richness estimates for these
clusters is in progress.

\subsection{\emph{ROSAT} All-Sky Survey}

The \emph{ROSAT} All-Sky Survey~\citep[RASS, ][]{vabbb99} took place primarily
during a six-month campaign in 1990-1991 to image the whole sky in soft X-rays
(0.1-2.4 keV) with the ROSAT Position Sensitive Proportional
Counter~\citep[PSPC, ][]{pbhkm87}.  The survey scanned the sky in great
circles, with the largest net exposure time ($\sim 40000\,\mathrm{s}$) near the
ecliptic poles.  The typical field coincident with the maxBCG survey region,
which does not overlap the northern ecliptic pole, has an effective exposure
time of $\sim 400\,\mathrm{s}$.  The point spread function (PSF) for these RASS
scans is very broad (full-width-half-maximum of $\sim 25\arcsec$) and is
dominated by far off-axis photons due to the survey strategy.  \citet{vbefs01}
released reprocessed photon maps and exposure maps of the entire RASS survey
region.  These photon maps provide the input for the analysis described in this
paper.

RASS data has been used to create several catalogs of purely X-ray selected
objects.  The ROSAT bright
source catalog~\citep[BSC,][]{vabbb99} consists of 18,811 sources with a
typical signal-to-noise $>4$.  The position resolution is superior to the PSF
FWHM with 68\% (90\%) of the sources within $13\arcsec$ ($25\arcsec$).  In
addition to the BSC, there is a companion Faint Source
Catalog~\citep[FSC,][]{vabbb00} consisting of 105,924 sources with a typical
signal-to-noise $>2$. Most are unresolved.
Combined, these catalogs provide soft X-ray detected
sources which can also be compared to the maxBCG clusters.

The RASS photon data has also been used as an input for X-ray flux-limited
cluster catalogs.  The Brightest Cluster Sample~\citep[BCS,][]{eebac98} is a
flux-limited sample of the brightest 201 clusters in the northern hemisphere
with fluxes $F_X >
5\times10^{-12}\,\mathrm{erg}\,\mathrm{s}^{-1}\,\mathrm{cm}^{-2}$. The Northern
ROSAT All-Sky galaxy cluster survey~\citep[NORAS,][]{bvhmg00} is a catalog of
378 extended X-ray sources that have been confirmed to be clusters via optical
follow-up.  Due to the broad ROSAT PSF, this catalog is only $\sim50\%$
complete at their stated flux limit, $F_X >
3\times10^{-12}\,\mathrm{erg}\,\mathrm{s}^{-1}\,\mathrm{cm}^{-2}$, which
corresponds to a luminosity $L_X \gtrsim 1\times10^{44}\,\lum$ at our median
redshift $\zmed=0.23$.  \citet{kmawe07b} have performed an initial comparison
between NORAS clusters and maxBCG clusters and have found that the maxBCG
detects $>90\%$ of NORAS objects.  This is consistent with estimates of the
completeness from simulations~\citep{kmawe07b, rwkme07}.  The \emph{ROSAT}-ESO
Flux Limited X-ray galaxy cluster survey~\citep[REFLEX:][]{bsgcv04} is similar
to NORAS in the southern sky, with the same flux limit.  This catalog of 447
clusters is over 90\% complete due to improvements in RASS analysis, although
only a small fraction of REFLEX overlaps the maxBCG survey area.

\section{X-ray Analysis} \label{sec: x-ray analysis}

The typical RASS exposure time for maxBCG clusters, 400 s, is too short to
allow significant detections for individual clusters. The large number of
maxBCG clusters, however, allows us to make up for this. For example, there are
7986 clusters with richnesses 9 $\le N_{200} \le$ 11. For these objects, the
total RASS exposure time is $\approx 320,000\,\mathrm{s}$. Such a large total
exposure allows us to measure the mean X-ray emission from these clusters quite
precisely.

Details of the stacking method are described later in
this section, and outlined here.  We begin by dividing the clusters into 
nine richness ($N_{200}$) bins. To simplify comparison to other maxBCG 
analyses, we use the same richness bins used in measurements of the mean
velocity dispersions~(B07).  The number of clusters in each 
richness bin is shown in Table~\ref{tab:ncluster}.

\begin{deluxetable}{cccccc}
\tablewidth{0pt}
\tablecaption{\label{tab:ncluster}Number of Clusters In Each Bin}
\tablehead{
\colhead{Richness Range\tablenotemark{a}} & \colhead{$\nmean$} &
\colhead{$\zmed$} & \colhead{$R_{200}$\tablenotemark{b}} & \colhead{$N_{\mathrm{clust}}$} &
\colhead{$N_{\mathrm{stack}}$}\\
 & & $(\kpc)$ & &
}
\startdata
$71 \leq N_{200} \leq 188$ & 92.85 & 0.21 & 1727 & 55 & 55\\
$51 \leq N_{200} \leq 70$ & 58.22 & 0.21 & 1469 & 146 & 140\\
$41 \leq N_{200} \leq 50$ & 44.67 & 0.21 & 1317 & 207 & 201\\
$33 \leq N_{200} \leq 40$ & 35.74 & 0.21 & 1201  & 356 & 339\\
$26 \leq N_{200} \leq 32$ & 28.57 & 0.22 & 1102  & 665 & 633\\
$21 \leq N_{200} \leq 25$ & 22.70 & 0.22 & 997  & 1128 & 1060\\
$18 \leq N_{200} \leq 20$ & 18.91 & 0.23 & 941  & 1141 & 1099\\
$12 \leq N_{200} \leq 17$ & 13.88 & 0.23 & 823  & 5651 & 5405\\
$9 \leq N_{200} \leq 11$ & 9.80 & 0.23 & 727 & 7986 & 7566\\
random & & & 7986 & 7529\\
\enddata
\tablenotetext{a}{$N_{200}$ is the number of red-sequence galaxies brighter
  than 0.4 $L_*$ within a scaled aperture $R_{200}$.}
\tablenotetext{b}{$R_{200}$ is the radius internal to which the mean density is
200 times the critical density~\citep{hmwas05}.}
\end{deluxetable}

As we do not have X-ray centers for individual clusters, we treat the BCG
selected by the maxBCG algorithm as the center of each cluster and stack on
these centers.  In \citet{kmawe07b} it was shown that most maxBCG-selected BCGs
($\sim80\%$) agree
well with the center of X-ray selected clusters, which has also been seen for
other optically-selected catalogs~\citep[e.g.][]{lm04}.  Possible biases 
introduced by
this assumption are discussed in Section~\ref{sec:centering}.  Every source and
background photon in the stacked analysis is scaled and weighted to the median
redshift of the clusters in the catalog, $\zmed=0.2296$.  With these weighted
photon maps we construct stacked images in Section~\ref{sec:images}, radial
profiles in Section~\ref{sec:profiles}, and spectra and luminosities in
Section~\ref{sec:spectra}.

\subsection{\label{sec:extraction}Cluster Extraction and Selection}

We obtain RASS photon data and merged exposure maps from the archives available
at the High Energy Astrophysics Science Archive Research Center (HEASARC).  The
RASS data and exposure maps are distributed in arbitrarily constructed $6.4
\times 6.4$ degree fields with significant overlap.  Nevertheless, since we
utilize a large background annulus extending to $40'$ from the central BCG, a
significant fraction ($\sim20\%$) of the maxBCG clusters fall across a field
boundary.  We have therefore built a tool that extracts all RASS photons in a
given aperture, from multiple RASS fields if necessary, counting only once
those photons which appear multiple times in overlap regions.\footnote{RASS
photons are tagged with the detector location, energy channel, and time of
arrival, providing a unique description of each photon.}  After photon
retrieval, the appropriate merged exposure maps ({\tt mex} file) are used to
calculate the effective exposure time ($t_i$) at the position of each detected
photon.

Before stacking we exclude from the list a subset of maxBCG clusters which
might bias our X-ray measurements for unphysical reasons.
No cut on Galactic absorption is required, as the SDSS observations
are restricted to high galactic latitude.  The typical equivalent Galactic
hydrogen column density is a very low: $N_H =
2\times10^{20}\,\mathrm{cm}^{-2}$~\citep{dl90} for the maxBCG clusters, with a
maximum of $N_H = 9\times10^{20}\,\mathrm{cm}^{-2}$.  Although the typical RASS
exposure time is $\sim400\,\mathrm{s}$, there are some
fields with significantly less.  This makes source and background
estimation difficult, while not adding significantly to the signal.
For this reason, we reject all clusters with less than 
200~s mean exposure time.
This removes 4\% of the total number of clusters, but less than 1\% of the net
exposure time.

An important possible source of contamination for stacked cluster measurements
is the presence of a few extremely bright foreground sources. We would like to
reject regions contaminated by these sources. RASS images of maxBCG clusters,
especially at higher redshift, are often unresolved, making it difficult to
remove point sources using extent information, so we proceed as follows. We
first identify those ROSAT Bright Source Catalog (BSC) sources in the survey
area with count rates higher than that expected for emission from any cluster
at redshift beyond 0.1. This corresponds to a count rate of $\approx
2\,\mathrm{ct}\,\mathrm{s}^{-1}$. There are 179 BSC sources in the maxBCG
survey area with ROSAT soft-band (0.1-0.5 keV) or hard-band (0.5-2.1 keV) count
rates above this limit.  Of these 179 sources, only 13 are clusters as
identified in the \emph{ROSAT} Brightest Cluster
Sample~\citep[BCS,][]{eebac98}.  Of these 13 clusters, only one of these
objects is associated with a maxBCG cluster (Abell 2142, the richest and one of
the nearest clusters in the catalog). The other high-flux clusters are at
redshifts $z \ll 0.1$.  Further visual inspection has confirmed that Abell 2142
is the only maxBCG cluster associated with one of these extremely bright
sources.  For the stacking analysis, we reject all maxBCG clusters (except for
Abell 2142) within $45\arcmin$ of any of these bright sources, due to possible
contamination from non-cluster photons.  This removes only 0.6\% of the maxBCG
catalog. 

The Virgo cluster is another important foreground source.  This nearby, X-ray
bright cluster dominates the RASS emission over a significant region of sky,
thus making it difficult to estimate the X-ray emission from background
clusters.  We therefore reject all clusters that fall
within $1.5$ degrees of M86 or M87.  This is a relatively small 
effect, removing 
only 0.1\% of the maxBCG clusters.  Table~\ref{tab:ncluster}
shows the number of clusters in each richness bin before and after the
rejections outlined in this section, as well as the mean richness for each bin.

To stack the X-ray measurements of the clusters, RASS photons from each cluster
must be scaled and appropriately weighted. The projected physical distance to
the BCG is calculated for each photon, and this scaled distance is used in
image construction and radial profile calculations. Each source and background
photon is then weighted by a factor of $w(z) = (d_L(z)/d_L(\zmed))^2$, where
$d_L(z)$ is the luminosity distance to the cluster and $d_L(\zmed)$ is the
luminosity distance to the median cluster redshift $\zmed=0.2296$. For
background estimation we use a fixed annulus with inner (outer) radius of
$20\arcmin$ ($40\arcmin$), similar to that used for the NORAS catalog for
individual clusters~\citep{bvhmg00}.  This background annulus corresponds to a
physical distance of $1.5 - 3.1\,\Mpc$ at our minimum redshift of $0.1$ and a
$3.8 - 7.5\,\Mpc$ at our maximum redshift of $0.3$.  We confirm that our
results are not sensitive to the exact choice of background annulus.  To
calculate the weighted count rate ($C_w$) in a given bin (image, radial, or
spectral) for a cluster at redshift $z$, we follow a procedure similar to
\citet{bvhmg00}, where $C_w = \sum_i w(z_i)/t_i$, where $t_i$ is the exposure
time at the position of each detected photon.

As a simple test of our stacking procedure, we also extract RASS photons
from a set of positions selected at random from the region covered by the 
maxBCG catalog. Each such random position is associated with a real cluster
redshift drawn from the catalog in the $9 \leq N_{200} \leq 11$ bin.  
The random points are analyzed identically to the real cluster positions.

\subsection{\label{sec:images}Stacked X-Ray Images}

\begin{figure*}
\begin{center}
\plotone{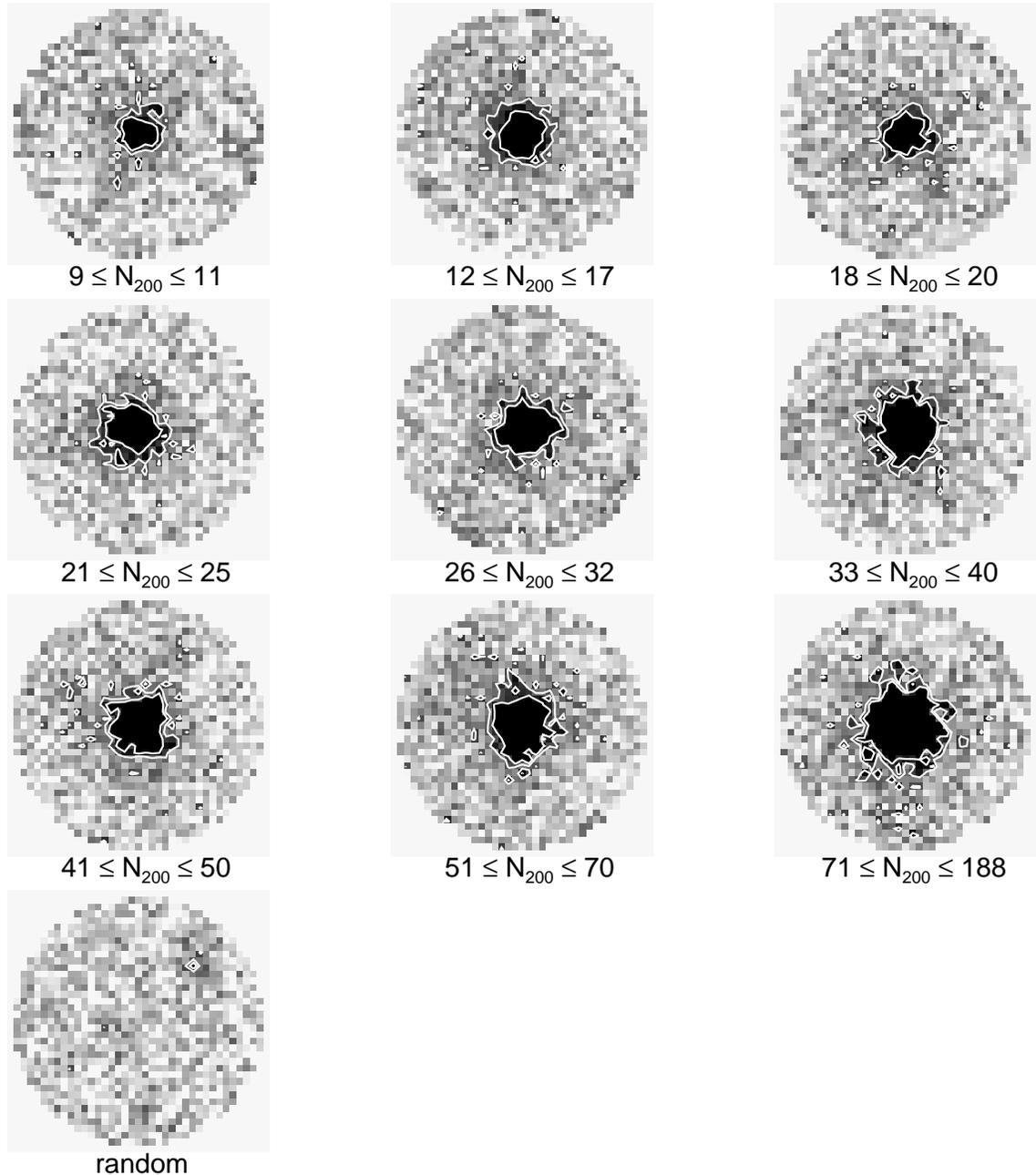}
\caption{\label{fig:imageplots}Stacked X-ray images for the 9 richness bins and
random points.  Each image has a projected radius of $2.0\,\Mpc$.  The
\emph{ROSAT} hard band (0.5-2.0 keV) counts have been background subtracted and
put in $100\times100\,\kpc$ bins.  For display, each image has been scaled with
histogram equalization to show similar background noise levels.  The contours
are drawn $3\sigma$ and $5\sigma$ above background level.}
\end{center}
\end{figure*}

The background-subtracted stacked images are shown in
Figure~\ref{fig:imageplots}.  Each image has a projected radius of
$2.0\,\Mpc$.  The images contain photons from the \emph{ROSAT} hard
band (channels 51-201, 0.5-2.0 keV) as this band has the highest
signal-to-noise for cluster emission.
After background subtraction, the counts were put in
$100\times100\,\kpc$ bins.  For display, each image has been
scaled with histogram equalization to show similar background noise levels, as
there are two orders of magnitude more clusters in the poorest bin than the
richest bin.  The contours are drawn $3\sigma$ and $5\sigma$ above background
level.

Highly significant X-ray emission is seen in each stack of clusters.  By
contrast, in a stack of 7,566 random points there is no significant excess, and
no sign of emission centered in the stack. It is worth noting that the average
count rate in the $9 \leq N_{200} \leq 11$ bin is only
$0.0025\,\mathrm{ct}\,\mathrm{s}^{-1}$, corresponding to a flux $F_X =
5\times10^{-14}\,\mathrm{erg}\,\mathrm{s}^{-1}\mathrm{cm}^{-2}$.
There are a total of $\sim 8500\,\mathrm{cts}$ in excess of background from 
$\sim 8000$ clusters; an average of only 1 source count per cluster.  This 
illustrates the power of the stacking exercise to probe very low count rates, 
and allows us to use RASS to measure very low $L_X$ cluster emission 
even at a moderate redshift of $\sim 0.2$.

\subsection{\label{sec:profiles}Radial Profiles}

\begin{figure*}
\begin{center}
\scalebox{0.7}{\rotatebox{270}{\plotone{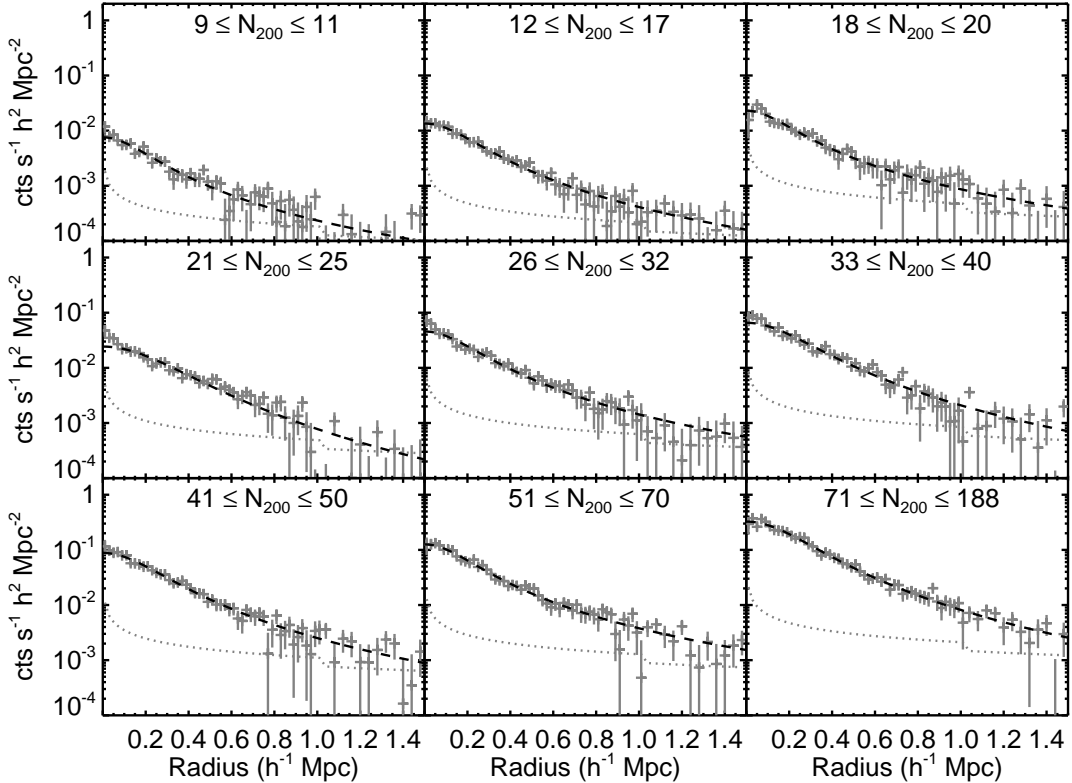}}}
\caption{\label{fig:profiles}The background-subtracted surface-brightness
  profiles for the stacked images in each richness bin.  The radial profiles
  have been summed in $50\,\kpc$ ($100\,\kpc$) bins inside (outside) $1\,\Mpc$.
  For reference, the dotted lines represent the $1\sigma$ background limit.
  Significant cluster emission is seen in each richness bin.  For the richest
  bins, this extends out beyond $1\,\Mpc$. The dashed lines show the $\beta$
  model fits from Table~\ref{tab:betafits}.}
\end{center}
\end{figure*}

Figure~\ref{fig:profiles} presents the background-subtracted surface-brightness
profiles for the stacked X-ray images in each richness bin. The radial profiles
have been summed in $50\,\kpc$ ($100\,\kpc$) bins inside (outside) $1\,\Mpc$.
For reference, the dotted lines represent the $1\sigma$ background level.
Significant cluster emission is seen in each richness bin.  For the richest
bins, this extends out beyond $1\,\Mpc$. The X-ray surface brightness profiles
for the different bins look remarkably similar, except for the normalization
which increases strongly with richness and the signal-to-noise which decreases
slowly with richness.  We fit the surface brightness profiles out to
$1.5\,\Mpc$ with a standard $\beta$ model, $S(R) =
S_0(1+R^2/R_C^2)^{-3\beta+1/2}$.  The best-fit parameters are presented in
Table~\ref{tab:betafits}, and are overplotted with dashed lines in
Figure~\ref{fig:profiles}.  The $\beta$ model results in a good fit in all
cases.

\begin{deluxetable}{cccc}
\tablewidth{0pt}
\tablecaption{\label{tab:betafits}$\beta$ Model Parameters}
\tablehead{
\colhead{Richness Range} & \colhead{$\beta$} & \colhead{$R_c (\Mpc)$} &
\colhead{$\chi^2/\nu$}
}
\startdata
$71 \leq N_{200} \leq 188$ & $0.67\pm0.03$ & $0.31\pm0.03$ & 37.7/72\\
$51 \leq N_{200} \leq 70$ & $0.55\pm0.03$ & $0.23\pm0.02$ & 43.2/72\\
$41 \leq N_{200} \leq 50$ & $0.62\pm0.04$ & $0.28\pm0.03$ & 36.2/72\\
$33 \leq N_{200} \leq 40$ & $0.64\pm0.04$ & $0.31\pm0.03$ & 75.1/72\\
$26 \leq N_{200} \leq 32$ & $0.57\pm0.03$ & $0.24\pm0.03$ & 46.3/72\\
$21 \leq N_{200} \leq 25$ & $0.75\pm0.08$ & $0.40\pm0.05$ & 88.1/72\\
$18 \leq N_{200} \leq 20$ & $0.51\pm0.03$ & $0.21\pm0.03$ & 38.9/72\\
$12 \leq N_{200} \leq 17$ & $0.57\pm0.03$ & $0.24\pm0.03$ & 36.2/72\\
$9 \leq N_{200} \leq 11$ & $0.54\pm0.04$ & $0.22\pm0.04$ & 63.4/72\\
\enddata
\end{deluxetable}

The core radii, $R_c$, for these $\beta$ model fits are surprisingly large,
ranging from $\sim200\,\kpc$ to $\sim400\,\kpc$.  These are significantly
larger than those typically seen for X-ray clusters~\citep[$\sim50-150\,\kpc$,
e.g.][]{na99}.  The $\beta$ parameters are typical for X-ray
clusters~\citep[e.g.][]{na99, spflm03}. There is a slightly significant ($\sim
3\sigma$) trend in $\beta$ with richness, such that the richer clusters tend to
have slightly larger values of $\beta$.  This slight trend is in the same sense
as has been seen in previous work correlating $\beta$ with cluster
mass~\citep{hms99, hp00, spflm03}.  These $\beta$-model parameters are similar
to those measured by DKM07, in which $\beta$ was not seen to be strongly
correlated with richness.  These inconclusive results are caused by two
effects; the broad point spread function for RASS objects, and the offset
distribution between the BCG and the X-ray cluster emission.

\subsubsection{\label{sec:rasspsf}RASS PSF effects}

\begin{figure}
\begin{center}
\scalebox{0.85}{\rotatebox{270}{\plotone{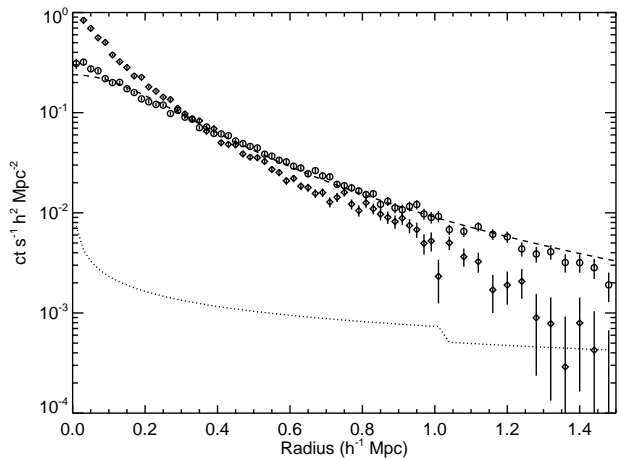}}}
\caption{\label{fig:ptsrc}The radial profile of stacked point sources selected
  from the WGA~\citep{wga00} catalog.  Each point source has been randomly
  assigned a redshift from the maxBCG redshift distribution to simulate the
  effect of smearing due to radial scaling.  The diamonds show the stacked
  point source profile with the centers as determined in WGA; there is
  significant emission out to distances $> 1\,\Mpc$ ($7\arcmin$ at
  $\zmed=0.22$).  The circles show the stacked point source profile after
  convolving with the estimated maxBCG/BSC-FSC offset distribution.  For
  reference, the dotted line represents the $1\sigma$ background limit.  The
  outer profile is well-fit by a $\beta$-model (dashed line) with
  $\beta=0.61\pm0.01$.  Thus, a cluster-like radial profile is obtainable in
  stacked images even without true extended cluster emission.  }
\end{center}
\end{figure}

To study the effect of the RASS PSF on the stacked profiles, we test our
analysis on bright ROSAT point sources.  These sources were selected from the
WGA catalog~\citep{wga00}, which contains 88,621 well measured point-like
sources selected from ROSAT PSPC pointed observations.  We take $\sim 500$
moderately bright ($0.1 - 0.5\,\mathrm{ct}\,\mathrm{s}^{-1}$) point sources
that overlap the SDSS DR5 mask.  These bright sources are nearby Galactic
sources as well as AGN and quasars at moderate and high redshift, and are also
well detected in RASS.  Each point source is randomly assigned a redshift drawn
from the maxBCG cluster catalog redshift distribution.  We note that a nearby
unresolved point source is indistinguishable from a distant unresolved point
source.  The point sources are then stacked in exactly the manner described for
clusters above.  Figure~\ref{fig:ptsrc} shows the radial profile of the stacked
point sources (diamonds), which is essentially a measure of the RASS PSF with
the radial scaling calculated for the maxBCG cluster redshifts.  While this
radial profile is much more sharply peaked than that of the stacked clusters,
the broad ROSAT PSF scatters significant X-ray emission to distances $>
1\,\Mpc$ ($7\arcmin$ at $z=0.2$) from the center.  This PSF smearing implies an
effective minimum on the $\beta$-model core radius, as well as creating a
possible bias in the calculation of the $\beta$ parameter.  We note that
$\lesssim 10\%$ of the selected WGA point sources are within $10\arcmin$ of
maxBCG clusters.  Excluding the point sources that are neighboring these
possibly extended X-ray sources does not significantly alter this result.

\subsubsection{Optical / X-ray offsets}

The stacked, projected surface brightness profile may also
be affected by our choice
of cluster centers.  There are several ways in which the BCG chosen by the
maxBCG algorithm might be offset from the X-ray emission from the cluster.
Dynamically active clusters and clusters in the process of merging can have a
significant offset between the X-ray centroid and the BCG.  There is also the
possibility that the maxBCG algorithm chose an incorrect center (this is
addressed more fully in Section~\ref{sec:centering}).  In addition, there may
be additional X-ray point sources such as AGN that are associated with the
clusters.  All of these effects result in an effective optical/X-ray offset
distribution that may bias the radial profiles.

\begin{figure}
\begin{center}
\scalebox{0.85}{\rotatebox{270}{\plotone{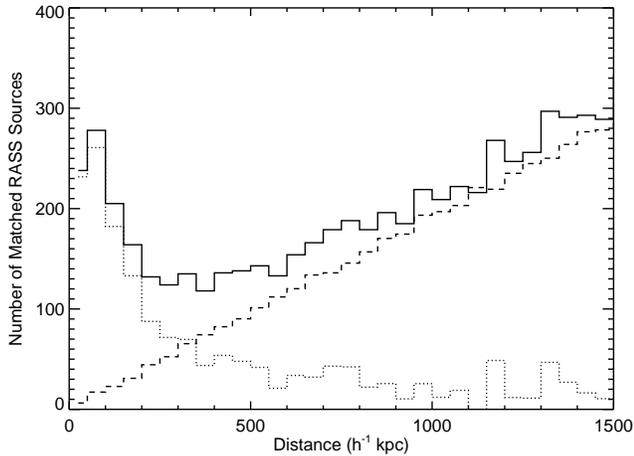}}}
\caption{\label{fig:ptmatchhist}Histogram of offset distribution between maxBCG
  clusters and X-ray sources detected in the BSC and FSC catalogs.  Each
  cluster is allowed to match to multiple X-ray sources, ensuring we find all
  associated X-ray sources in addition to all random matches.  The solid
  (dashed) line shows the histogram of maxBCG (random) offsets in $50\,\kpc$
  bins.  The dotted line shows the result of subtracting the random matches
  from the cluster matches.  There is a tight core in which the BCG is within
  $\sim150\,\kpc$ of an X-ray source, as well as a long tail.}
\end{center}
\end{figure}

We model the optical/X-ray offset distribution by matching the maxBCG catalog
to known X-ray sources.  For this exercise, we use the fact that many maxBCG
clusters ($>900$) are associated with individual detections in the ROSAT Bright
Source and Faint Source Catalogs.  Most of these BSC and FSC sources have not
been previously recognized as associated with clusters, primarily because they
are too faint or too distant to be seen as significantly extended in RASS.  We
match the maxBCG clusters to the BSC and FSC catalogs, allowing multiple X-ray
sources to match to each cluster.  This ensures that we find all X-ray sources
associated with each cluster, as well as all possible random matches.  We
similarly match an equal number of random locations drawn from the maxBCG
survey region to the BSC and FSC catalogs.  Figure~\ref{fig:ptmatchhist} shows
the projected offset distribution from the maxBCG clusters to the BSC and FSC
catalogs.  The solid (dashed) line shows a histogram of maxBCG (random) offsets
in $50\,\kpc$ bins.  The dotted line shows the residual distribution after
subtracting the random matches from the cluster matches.

The subtracted histogram in Figure~\ref{fig:ptmatchhist} shows a large excess
of X-ray sources associated with the optical cluster centers.  There is a tight core in
which the BCG is within $\sim150\,\kpc$ of an X-ray source, as well as a long
tail extending out to $\sim1500\,\kpc$.  The X-ray source excess at large
radius is likely to be associated with the maxBCG clusters.  These sources
comprise a mix of merging clusters, clusters with poorly identified centers, 
and associated point sources such as AGN.  The subtracted histogram is used as 
a first-order empirical radial distribution of X-ray sources associated with
maxBCG cluster centers.

We now estimate the additional effect on the radial profile due to the radial
offset distribution of X-ray sources associated with the maxBCG clusters.  We
take each WGA point source from Section~\ref{sec:rasspsf} and randomly alter
its position so that the distribution is offsets is identical to the empirical
radial distribution.  Although we have only measured the offset
distribution for the brightest $\sim5\%$ of maxBCG clusters, we assume this is
a good proxy for the offset distribution between the X-ray emission and BCGs of
all the maxBCG clusters.  These offset point sources are run through our stacking
analysis.  Figure~\ref{fig:ptsrc} shows the radial profile of the offset point
sources after stacking (circles).  This radial profile is reasonably well fit
with a $\beta$ model, although the fit is poor within $\sim 200\,\kpc$.  At
large radii the radial profile is well fit with $\beta = 0.61\pm0.01$.  Thus,
at large radii it is not possible to distinguish between true \emph{extended}
cluster emission and X-ray \emph{point} sources convolved with the observed
maxBCG--X-ray source offset distribution.

This calls into question the utility of the $\beta$ model parameterization for
this exercise.  The stacked X-ray profile is a convolution of the ROSAT PSF,
the centering distribution, and the true extended X-ray emission.
Operationally, it is not possible to separate the contributions from these
three components.  Therefore, the $\beta$ parameters that are the result of a
stacking exercise such as this one, where most of the individual X-ray clusters
are not detected, should be used cautiously.

\subsection{\label{sec:spectra}Stacked X-ray Spectra}

We perform spectral fitting of the photons in each stack to calculate X-ray
luminosities.  The spectra were analyzed using XSPEC version
11.3.2~\citep{a96}.  Spectral analysis on the stacked ROSAT data involves a
variety of complications, some of which are discussed in DKM07.  We outline
some key issues here.

All RASS observations integrate exposure time across the \emph{ROSAT}
field-of-view. As a result, we use the spectral response file {\tt
pspcc\_gain1\_256.rmf} from HEASARC, suitable for off-axis observations.  When
combined with the appropriate ancillary response file ({\tt arf}), we obtain
the vignetting corrected spectral response for the off-axis photons that
comprise our spectra.  However, the exposure times extracted from the merged
exposure maps for RASS are already corrected for vignetting, using the mean
spectrum of the X-ray background in the PSPC band\footnote{see the manual page
for the FTOOL {\tt pcexpmap}}.  In practice, this means that our spectra, fit
with the standard {\tt rmf}+{\tt arf} combination, are corrected for vignetting
\emph{twice} -- once by using the exposure times obtained from the merged
exposure map, and once in the calculation of the {\tt arf} file.  By comparing
the on-axis response file {\tt rsp} from HEASARC with the combination of the
{\tt rmf}+{\tt arf} we can approximate the extra vignetting correction.  This
correction is a function of energy, and depends (weakly) on the spectrum of the
observed source.  By simulating cluster spectra of various temperatures, we
calculate the typical correction factor averaged across the ROSAT hard band
(0.5-2.0 keV) which is dominated by cluster flux.  This correction factor is
$\sim 1.53$.  Using this correction factor results in very good agreement
between our spectral analysis and the REFLEX count-rate to flux conversion
tables of \citet{bsgcv04}.  We estimate that applying this correction factor to
all flux and luminosity values determined from stacked spectra adds an
additional systematic error of $\sim 10\%$.

For our analysis, we use the {\tt rmf} file described above, combined with an
{\tt arf} output with the FTOOL {\tt pcarf}.  As noted in DKM07, each {\tt arf}
file is essentially identical, because each stacked cluster from RASS samples
photons from the entire ROSAT field of view.  Spectral files are then grouped
with a minimum of 50 counts per bin (after background subtraction) to ensure
valid results using $\chi^2$ statistical analysis.  Fits performed with XSPEC
were restricted to the 0.1-2.1 keV range.  The uncertainties in spectral fit
parameters are 90\% confidence errors, obtained by allowing all fit parameters
to vary simultaneously.

We obtain cluster spectra by summing all weighted cluster photons in both fixed
physical apertures and scaled apertures of the optically determined $R_{200}$
(see Section~\ref{sec:maxbcg}).  The fixed aperture of $750\,\kpc$ is chosen as
a reasonable fiducial value because this provides good signal-to-noise, and
because the radial profiles do not appear to change significantly with
richness.  The $R_{200}$ value for each bin was taken as the median $R_{200}$
of all the clusters in the richness bin; these values are reported in
Table~\ref{tab:lxr200}.  Background spectra are stacked with the same weights
as the source spectra, using the annuli defined in
Section~\ref{sec:extraction}.  As described previously, each photon is weighted
to the median redshift $\zmed = 0.2296$.  Unfortunately, $k$-correction of
individual photons is not possible, due to the fact that the detection channel
of the incident photons is most strongly dominated by the spectral response of
the ROSAT instrument.  The stacking procedure will therefore tend to ``smear
out'' the incident cluster spectra.  Simple simulations of X-ray spectra using
XSPEC show that this does not create a large effect on our best-fit spectral
values, as our redshift range of $0.1 < z < 0.3$ is not particularly broad.
Similarly, we do not make any corrections for possible redshift evolution in
X-ray luminosity; we will address this further in Section~\ref{sec:redshift}.

\begin{deluxetable*}{ccccccc}
\tablewidth{0pt}
\tablecaption{\label{tab:lxr200}Spectral Fits within $R_{200}$}
\tablehead{
\colhead{Richness Range} & \colhead{$\nmean$} & \colhead{$R_{200}$} &
\colhead{$N_H$} & \colhead{kT} & \colhead{$\lmean$} & \colhead{$\chi^2/\nu$}\\
 & & ($\Mpc$) & ($10^{20}\,\mathrm{cm}^{-2}$) & (keV) & 
($10^{42}\,\lum$) &
}
\startdata
$71 \leq N_{200} \leq 188$ & 92.85 & 1.73 & $2.5\pm0.4$ & $3.3^{+1.4}_{-0.9}$ & $228\pm33$ &
47.2/76\\
$51 \leq N_{200} \leq 70$ & 58.22 & 1.47 & $1.5\pm0.3$ & $3.4^{+2.1}_{-1.1}$ & $80.6\pm6.9$ &
61.5/77\\
$41 \leq N_{200} \leq 50$ & 44.67 & 1.32 & $2.8^{+0.7}_{-0.6}$ & $2.5^{+1.1}_{-0.6}$ &
$58.0\pm7.3$ & 87.4/65\\
$33 \leq N_{200} \leq 40$ & 35.74 & 1.20 & $2.8^{+0.6}_{-0.4}$ & $2.0^{+0.6}_{-0.4}$ &
$47.7\pm4.8$ & 100.6/95\\
$26 \leq N_{200} \leq 32$ & 29.57 & 1.10 & $3.4^{+0.9}_{-0.6}$ & $1.7^{+0.4}_{-0.3}$ &
$28.4\pm3.0$ & 62.9/95\\
$21 \leq N_{200} \leq 25$ & 22.70 & 0.99 & $3.2^{+0.9}_{-0.6}$ & $1.6^{+0.4}_{-0.2}$ &
$18.5\pm2.1$ & 113.2/97\\
$18 \leq N_{200} \leq 20$ & 18.91 & 0.94 & $1.5^{+0.4}_{-0.3}$ & $1.6^{+0.4}_{-0.2}$ &
$12.5\pm1.1$ & 77.3/88\\
$12 \leq N_{200} \leq 17$ & 13.88 & 0.82 & $2.0^{+0.4}_{-0.3}$ & $1.4^{+0.3}_{-0.1}$ &
$6.84\pm0.47$ & 182.3/134\\
$9 \leq N_{200} \leq 11$ & 9.80 & 0.73 & $2.0^{+0.7}_{-0.5}$ & $1.1^{+0.08}_{-0.06}$ &
$3.30\pm0.39$ & 127.8/112\\
\enddata
\end{deluxetable*}

\begin{deluxetable*}{ccccccc}
\tablewidth{0pt}
\tablecaption{\label{tab:lx750}Spectral Fits within $750\,\kpc$}
\tablehead{
\colhead{Richness Range} & \colhead{$\nmean$} & \colhead{$R_{200}$} &
\colhead{$N_H$} & \colhead{kT} & \colhead{$L_X$} & \colhead{$\chi^2/\nu$}\\
 & & ($\Mpc$) & ($10^{20}\,\mathrm{cm}^{-2}$) & (keV) & 
($10^{42}\,\lum$) &
}
\startdata
$71 \leq N_{200} \leq 188$ & 92.85 & 1.73 & $1.8\pm0.3$ & $3.1^{+1.1}_{-0.7}$ & 
$167\pm24$ & 40.5/61\\
$51 \leq N_{200} \leq 70$ & 58.22 & 1.47 & $1.6\pm0.3$ & $3.8^{+2.2}_{-1.1}$ & 
$57.3\pm4.9$ & 40.1/59\\
$41 \leq N_{200} \leq 50$ & 44.67 & 1.32 & $2.2\pm0.4$ & $2.6^{+1.1}_{-0.6}$ &
$45.6\pm5.6$ & 85.6/61\\
$33 \leq N_{200} \leq 40$ & 35.74 & 1.20 & $2.3\pm0.4$ & $2.5^{+0.8}_{-0.5}$ &
$38.8\pm3.9$ & 92.0/81\\
$26 \leq N_{200} \leq 32$ & 28.57 & 1.10 & $3.0\pm0.6$ & $1.8^{+0.5}_{-0.2}$ &
$23.4\pm2.3$ & 78.1/84\\
$21 \leq N_{200} \leq 25$ & 22.70 & 0.99 & $2.6\pm0.5$ & $1.7^{+0.4}_{-0.2}$ &
$16.4\pm1.8$ & 86.0/94\\
$18 \leq N_{200} \leq 20$ & 18.91 & 0.94 & $2.0\pm0.5$ & $1.5^{+0.4}_{-0.2}$ &
$10.9/pm0.9$ & 73.6/76\\
$12 \leq N_{200} \leq 17$ & 13.88 & 0.82 & $2.0\pm0.3$ & $1.4^{+0.3}_{-0.1}$ &
$6.50\pm0.45$ & 182.3/134\\
$9 \leq N_{200} \leq 11$ & 9.80 & 0.73 & $2.1\pm0.7$ & $1.1^{+0.2}_{-0.06}$ &
$3.40\pm0.42$ & 137.6/114\\
\enddata
\end{deluxetable*}

The spectra are fit with an absorbed thermal plasma model~\citep{rs77}.  The
metallicity is fixed at 0.3 solar~\citep[e.g.][]{ag89}, and the redshift is set
to the median scaled redshift of 0.2296.  The luminosities are calculated in
the rest-frame 0.1-2.4~keV band at the median redshift.  We prefer to calculate
the 0.1-2.4~keV luminosity rather than bolometric luminosity
($\lxbolo$), due to the large extrapolations required to obtain
bolometric values, which strongly depend on spectral temperatures (see below).
The best-fit spectral parameters for the scaled $R_{200}$ apertures are shown
in Table~\ref{tab:lxr200}, and the parameters for the fixed $750\,\kpc$
apertures are shown in Table~\ref{tab:lx750}.

In order to account properly for the variations in cluster luminosity in each
richness bin, we use bootstrap resampling to estimate the luminosity
errors.  In each richness bin, we run 2000 trials by sampling the same number
of clusters in that bin with replacement.  To save considerable time with
processing, we did not recreate the entire stacking procedure for each
resampling.  Instead, we take the individual cluster counts, scaled to the
median redshift.  We confirm that the average scaled count rate is a
good proxy for luminosity in the 0.1-2.4 keV band (see
Section~\ref{sec:scatter} for details on this calculation).  The 68\%
($1\sigma$) confidence interval obtained from the bootstrap is added in
quadrature with the $L_X$ errors obtained from the spectral fits.  Only the
richest ($71 \leq N_{200} \leq 188$) bin is dominated by the bootstrap error
calculation, due to the large range in richness and $L_X$ in the bin.

\begin{figure}
\begin{center}
\scalebox{0.85}{\rotatebox{270}{\plotone{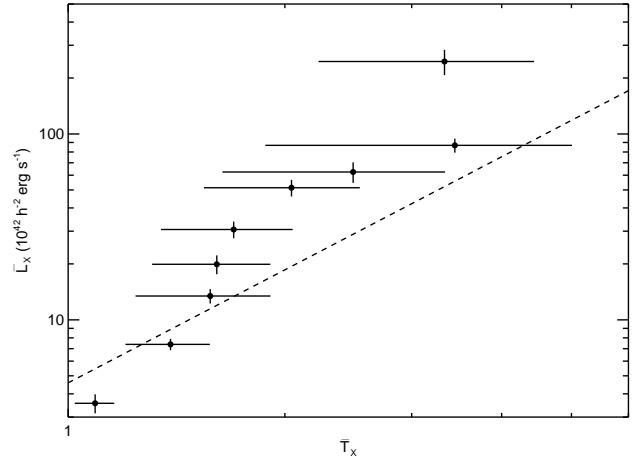}}}
\caption{\label{fig:ltplot}Mean $\lmean$--$\tmean$ relation for the stacked
  clusters, with the fits within $R_{200}$.  The dashed line shows the
  \citet{m98} relationship for ``uncorrected'' 0.1-2.4 keV luminosity.  The
  observed relation is much steeper than that derived by \citet{m98}.  This is
  due to the fact that \emph{ROSAT} temperatures tend to underestimate cluster
  temperatures with $T_X \gtrsim 3\,\mathrm{keV}$, as well as the complications
  in stacking many clusters with different temperatures.
}
\end{center}
\end{figure}

Figure~\ref{fig:ltplot} shows the mean $\lmean$--$\tmean$ relation for our
stacked clusters (within the scaled $R_{200}$ aperture), compared to the
\citet{m98} $L_X$--$T_X$ (0.1-2.4 keV) relation.  We see the cluster
temperature increase with luminosity, but our observed $\lmean$--$\tmean$
relation is noticeably steeper than that of \citet{m98}.  The stacked X-ray
temperatures appear to underestimate the expected temperature, especially at
$\lmean$ larger than a few times $10^{43}\,\lum$.  This discrepancy highlights
the challenge of measuring cluster temperatures with ROSAT, as well as the
challenges of measuring stacked cluster temperatures.  First, ROSAT has
sensitivity only to soft X-rays.  When the break of the bremsstrahlung
spectrum, determined by the temperature of the hot gas, is above $\sim
2.0\,\mathrm{keV}$, then the X-ray temperature becomes difficult to constrain.
The \emph{ROSAT} temperatures of hot clusters are generally underestimated; a
particular example is Abell 1689.  This bright cluster has a nearly isothermal
profile in \emph{XMM/Newton} observations with a temperature of $\sim
9\,\mathrm{keV}$ (consistent with the ASCA and \emph{Chandra} values), but has
a best-fit \emph{ROSAT}/PSPC temperature of
$4.3^{+1.2}_{-0.8}\,\mathrm{keV}$~\citep[e.g.][]{am04}.  Second, when we stack
many non-isothermal clusters with different temperatures, our isothermal
Raymond-Smith spectrum is no longer appropriate.  For example, \citet{rmbmd05}
have shown for \emph{Chandra} observations how the spectroscopically-weighted
temperature can differ significantly from the emission-weighted temperature for
single clusters, with the better measured colder gas dominating the spectral
fit.  When these effects are combined, interpreting the average best-fit
spectral temperature from a stack of hundreds of clusters is non-trivial.

Most importantly for our task of measuring the mean $\lmean$, the ROSAT
luminosity in the 0.1-2.4 keV band is virtually insensitive to the cluster
temperature.  If we fix the spectral temperature to the value predicted by the
\citet{m98} $L_X$--$T_X$ relation, then $\lmean$ changes by $\lesssim 5\%$ in
each bin, with no systematic bias with richness.  Similarly, we can follow the
prescription of \citet{bsgcv04} to convert the mean 0.5-2.0 keV RASS count rate
to flux and luminosity.  This method assumes the \citet{m98} $L_X$--$T_X$
relation, and requires iteratively calculating the flux and temperature.  The
$\lmean$ values thus obtained are consistent with those from the spectral
fitting to within $\lesssim 3\%$.  All this suggests that calculating
0.1-2.4~keV $L_X$ obtained from \emph{ROSAT}---which relies primarily on photon
counting---is more robust than calculating $T_X$ and extrapolating to
$\lxbolo$. Although a small $k$-correction ($\sim 10\%$) is required to
extrapolate from observer frame 0.1-2.4~keV to rest-frame luminosity, this
correction is not very sensitive to the spectral temperature at the moderate
redshift of the maxBCG clusters.

\section{Mean $\lmean$--$\nmean$ Relation}
\label{sec:meanrelation}

\begin{figure}
\begin{center}
\scalebox{0.85}{\rotatebox{270}{\plotone{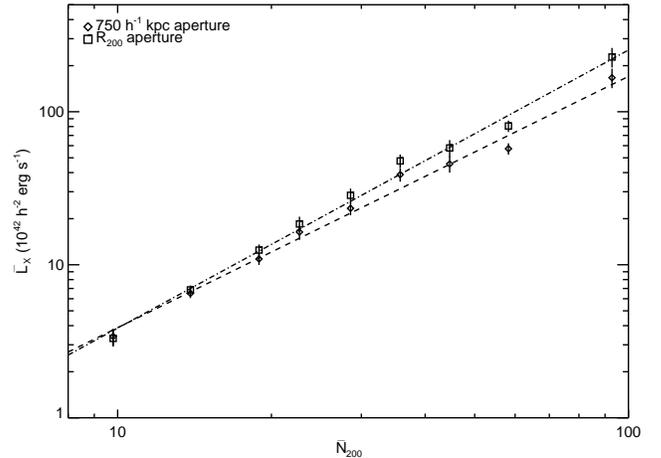}}}
\caption{\label{fig:lxn200plot}Mean $\lmean$--$\nmean$ relation for maxBCG
  clusters, for the fixed $750\,\kpc$ apertures (diamonds) and the scaled
  $R_{200}$ apertures (squares).  There is a strong correlation between
  $\lmean$ and $\nmean$, with a slope of $1.64\pm0.05$ ($750\,\kpc$) or
  $1.82\pm0.05$ ($R_{200}$).  The scaled aperture relation is steeper because
  the aperture scales with $N_{200}$.
}
\end{center}
\end{figure}

We stack the clusters and calculate the 0.1-2.4 keV X-ray luminosity as
described in the previous section.  Figure~\ref{fig:lxn200plot} shows the mean
$\lmean$--$\nmean$ relation, both for a fixed $750\,\kpc$ aperture and for the
scaled $R_{200}$ apertures.  We find a strong correlation between $\lmean$ and
$\nmean$ with both the fixed and the scaled apertures.  In the low richness
bins ($N_{200} \le 17$), the scaled aperture $R_{200}$ is approximately equal
to the fixed $750\,\kpc$ aperture, and $\lmean$ is approximately the same for
both choices of aperture.  In the high richness bins, the scaled aperture
$R_{200}$ is significantly larger than $750\,\kpc$, yielding a slightly larger
$\lmean$ for the scaled apertures. For this reason, the $\lmean$--$\nmean$ 
relation is slightly steeper using scaled apertures. The best-fit mean 
relationships are:

\begin{equation}\label{eqn:lxn_r200}
\lmean (<R_{200}) = e^{3.87\pm0.04} \left ( \frac{N_{200}}{40} \right
)^{1.82\pm0.05}\,10^{42}\,\lum
\end{equation}
\begin{equation}\label{eqn:lxn_750}
\lmean (<750\,\kpc) = e^{3.63\pm0.04} \left ( \frac{N_{200}}{40} 
\right )^{1.64\pm0.05}\,10^{42}\,\lum
\end{equation}

Power law fits are chosen to pivot around $N_{200} = 40$ to approximately
decouple errors in slope and normalization.

DKM07 also found that the mean X-ray luminosity of the 2MASS clusters scales
with optical richness ($N_{*666}$).  A direct comparison between the DKM07
relation and our relation is challenging, due to the different luminosity and
richness definitions we employ, but a first-order conversion is possible.
First, we convert the DKM07 bolometric luminosities (their Table 2) to
\emph{ROSAT} 0.1-2.4 keV luminosities using Table 5 of
\citet{bsgcv04}\footnote{This does not introduce any additional error, as DKM07
have published the precise spectral temperature they used to calculate
$\lxbolo$.}  Re-fitting for the $\lmean$--$N_{*666}$ relation, we
find $\lmean \propto (\bar{N}_{*666})^{1.51 \pm 0.07}$.\footnote{We note that
the soft band $\lmean$--$N_{*666}$ relation does not deviate from a power-law
at any richness, unlike the $\bar{L}_{X,\mathrm{bolo}}$--$N_{*666}$ relation in
Figure 10 of DKM07.}  As the 2MASS catalog is not yet public, we cannot make a
direct comparison of $N_{200}$ to $N_{*666}$ for individual clusters.
Futhermore, most of the clusters in the 2MASS catalog are at $z < 0.1$, while
the maxBCG catalog covers $0.1 \le z \le 0.3$.  At this time, the most we can
say is that the slope of the $\lmean$--richness relation is roughly similar
between our work and DKM07.

\subsection{\label{sec:scatter}Scatter in the $L_X$--$N_{200}$ Relation}

Comparisons to X-ray selected catalogs (see Section~\ref{sec:xrayselect}) as
well as prior studies~\citep[e.g.][]{seeb98,gbcz04} lead us to expect
significant scatter in the X-ray luminosity--richness relation for individual
clusters.  With a large scatter, the mean statistic used to calculate the
stacked $\lmean$-$\nmean$ relation may be significantly biased relative to the
median relation.  To understand this bias, 
we first assume that there is a log-normal conditional probability $p$
for the X-ray luminosity of a cluster at richness $N_{200}$:
\begin{equation}
p(L_{x}|N_{200}) = \frac{1}{\sqrt{2\pi}\siglnl}\exp \left [{\frac{-(\ln L_X - \lnlmean)^2}{2\siglnl^2}}\right ],
\end{equation}
with constant intrinsic scatter $\siglnl$, and mean log luminosity that follows:
\begin{equation}
\lnlmean(N_{200}) = A + B\ln (N_{200}/40),
\end{equation}
where $A$ is the log normalization of $L_X$ at $N_{200} = 40$, and $B$ is the
slope of the $L_X$--$N_{200}$ relation.  Note that $\exp(\lnlmean)$ for a
log-normal distribution is equivalent to the median (and geometric mean) of the
distribution.  For the duration of this paper, we employ the notation $\lmed
\equiv \exp(\lnlmean)$.

At a given richness, we wish to measure the median X-ray luminosity $\lmed$,
the peak of the underlying log-normal distribution.  However, the stacking
exercise we have undertaken is fundamentally a calculation of the
\emph{arithmetic mean} ($\lmean$) at a given $\nmean$.  For a log-normal
distribution with median $\lmed$ and intrinsic scatter $\siglnl$, the
arithmetic mean is $\exp(\lnlmean+\sigma_{\ln L}^2/2)$.  Thus, the stacked
normalization is an overestimate of the median of the underlying distribution
by a factor of $\exp(\siglnl^2/2)$.  If the scatter is large, the stacked
(mean) normalization will be dominated by the most luminous clusters and will
be biased high.  For example, an $80\%$ scatter indicates a $\sim30\%$ bias.

To constrain this scatter, we begin with measurements of X-ray flux (and hence
$L_X$) at the locations of each cluster with $N_{200} \ge 30$.  Though many of
these are low signal-to-noise detections, they can be used to measure
scatter. The IDL Astronomy library tool {\tt linmix\_err} is used to fit $L_X$
as a function of $N_{200}$ with intrinsic scatter $\siglnl$~\citep{k07}.  This
tool uses a Bayesian approach to linear regression with errors in X and Y and
is well behaved even when the measurement errors dominate.  It also handles
non-detections and upper limits in Y.  Monte Carlo simulations show that
the selected $N_{200}$ cut provides a large enough richness range to constrain
the slope and scatter, and provides larger signal-to-noise than the entire
cluster catalog, as over $80\%$ of the 955 clusters with $N_{200} > 30$ are
detected at at least the $1\sigma$ level.  Furthermore, the fits are not
sensitive to the precise richness cutoff chosen.

To calculate the $L_X$ for each of these often marginally-detected clusters, we
use a procedure based on the count-rate to flux conversion method from the
REFLEX survey~\citep{bsgcv04}.  First, we calculate the 0.5-2.0 keV (ROSAT hard
channel) count rate in a $750\,\kpc$ aperture; the fixed aperture was used
because it results in better signal-to-noise than a larger scaled aperture
(where we only see significant signal in the stacked profiles).  The local
background is calculated in a $20\arcmin-40\arcmin$ annulus using the
sector-rejection method of \citet{bvhmg00}.

In REFLEX the temperature is estimated in an iterative fashion from the
luminosity, using the $L_X$--$T_X$ relation of \citet{m98}.  As most of the
maxBCG clusters do not have a significant flux (or luminosity) measurement, we
cannot calculate the individual cluster temperatures in this way.  Instead,
each cluster temperature is approximated by the stacked spectral temperature
from the appropriate richness bin, as shown in Table~\ref{tab:lx750}.  This
ensures that our stacked cluster luminosities and individual luminosity
estimates are on approximately the same footing. Perhaps more importantly,
changing the temperature does not change the 0.1-2.4 keV luminosities
significantly~\citep[e.g.][]{mme99}.  For example, for the clusters with
$>1\sigma$ detections, if we follow the REFLEX iterative recipe to convert
count rate to $L_X$, rather than fixing the temperatures at the stacked values,
the individual $L_X$ values change by $<3\%$. This is simply another example of
how measuring temperatures with ROSAT is challenging.

The 0.5-2.0 keV count rate is then converted to 0.1-2.4 keV luminosity using
Table 2 from \citet{bsgcv04} with the equivalent hydrogen
column density at the position of the cluster~\citep{dl90}.  After converting
to luminosity, a $k$-correction is applied using Equation 4 from
\citet{sebsn06}, which is a good approximation of Table 3 in
\citet{bsgcv04}.\footnote{Note that there is a typographical error in Equation
4 of \citet{sebsn06}, so that it should read
$k(z,T)=\{1+[1+\log_{10}(T/5\,\mathrm{keV})]z\}^{1/2}$.}  The $k$-corrections
are not very large, at most 10\%.

\begin{figure*}
\begin{center}
\scalebox{0.7}{\rotatebox{270}{\plotone{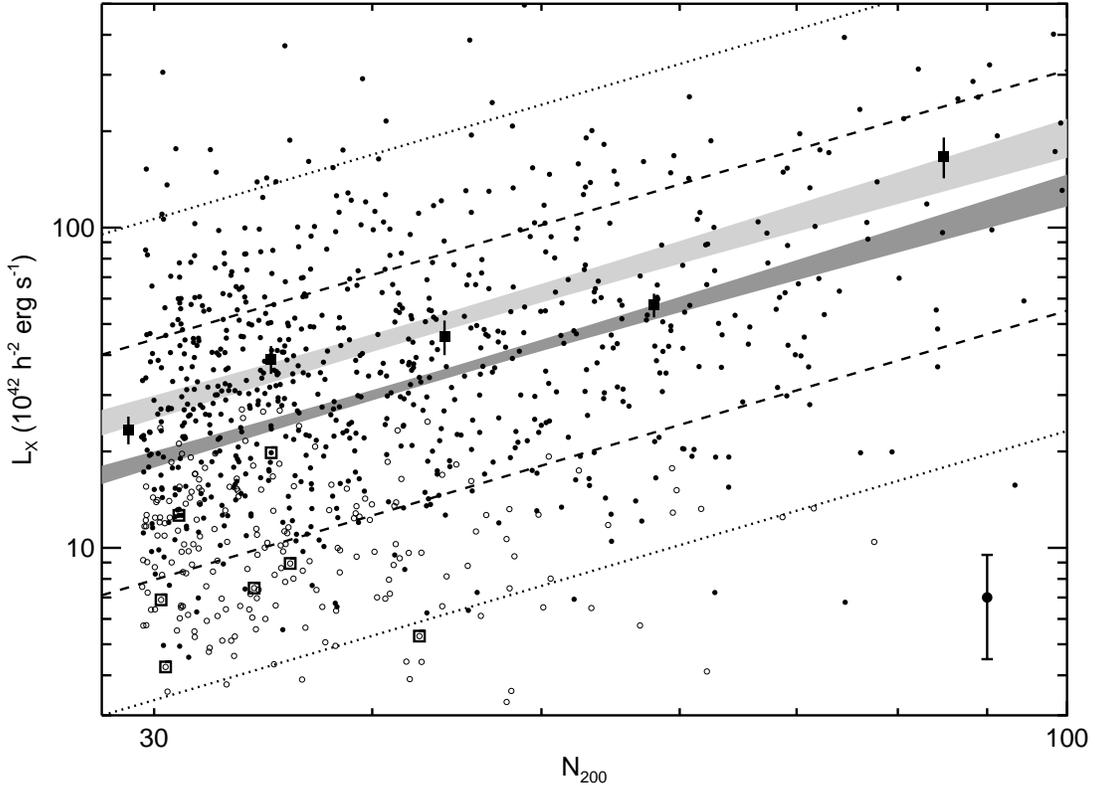}}}
\caption{\label{fig:linmix}$L_X$ vs. $N_{200}$ for individual clusters with
  $N_{200} \geq 30$.  The solid circles represent detections at the $1\sigma$
  level, and empty circles represent $1\sigma$ upper limits. The typical error
  bar is plotted on the fictitious data point in the lower-right
  corner. Contours showing the $\pm1\sigma$ contours on the best-fit median
  relation are shown in dark gray.  The dashed (dotted) lines show the
  $\pm1\siglnl$ ($\pm2\siglnl$) scatter constraints.  The median relation as
  constrained by {\tt linmix\_err} has been converted to the equivalent mean
  relation by multiplying the normalization by $\exp(\siglnl^2/2)$, and is
  shown in light gray.  The mean relation agrees well with the individual
  stacked bins (solid squares), which are about $\sim30\%$ brighter than the
  underlying median values.  The empty squares denote ``Abell X-Ray
  Underluminous'' (AXU) clusters~\citep{pbbr07}, which have X-ray luminosities
  consistent with the observed scatter.  }
\end{center}
\end{figure*}

We compared our method of calculating $L_X$ for individual clusters to the
$L_X$ values obtained in the NORAS catalog (see Section~\ref{sec:xrayselect} for
details on the X-ray catalog matching).  For the matched clusters the values
are all consistent within errors with $\sim10\%$ scatter, and a systematic
offset of $<5\%$. The primary difference in our calculations is that the NORAS
fluxes were calculated in an aperture obtained via growth curve analysis (GCA)
designed to obtain the best signal-to-noise for each cluster, while we use a
fixed physical aperture.  Therefore, our fixed temperature and aperture provide
an unbiased estimate of the cluster flux and luminosity even without \emph{a
  priori} knowledge of extended cluster emission. Individual cluster $L_X$ and 
$N_{200}$ values are shown in Figure \ref{fig:linmix}.

We use {\tt linmix\_err} to estimate the power-law slope, normalization, and
intrinsic scatter of the underlying distribution of the $L_X$--$N_{200}$
relationship for the 955 richest clusters with $N_{200} \geq 30$.  The best fit
relation is:
\begin{equation}
\lmed(<750\,\kpc) = e^{3.40\pm0.04} \left ( \frac{N_{200}}{40} \right )^{1.61\pm0.13}
\,10^{42}\,\lum,
\end{equation}
with an intrinsic scatter $\siglnl = 0.86\pm0.03$.  The individual $L_X$
values are calculated within a fixed $750\,\kpc$ aperture, and thus this is to
be compared to Equation~\ref{eqn:lxn_750}.  We test the robustness of
this result by both splitting the input data into multiple independently fit
subsamples and changing the richness threshold slightly.  We find that the
constraint on the scatter is robust, and the error bar is accurate.  Possible
systematic biases in the constraint on $\siglnl$ are discussed in
Section~\ref{sec:coolcore}. 

The scatter-corrected $\lmed$--$\nmean$ relation is shown in
Figure~\ref{fig:linmix}.  The luminosities for the individual $1\sigma$
detections are plotted as solid circles, and the $1\sigma$ upper limits are
plotted as empty circles.  The typical error bar for detections is shown in the
lower-right corner.  Contours showing the $\pm1\sigma$ contours on the best-fit
median relation are shown in dark gray, and the dashed (dotted) lines show the
$\pm1\siglnl$ ($\pm2\siglnl$) scatter constraints.  The median relation as
constrained by {\tt linmix\_err} has been converted to the equivalent mean
relation by multiplying the normalization by $\exp(\siglnl^2/2)$, and is shown
in light gray.  The mean relation agrees well with the individual stacked bins
(squares), which are about $\sim30\%$ brighter than the underlying median
values.

We must emphasize that we are measuring the \emph{observed} scatter in the
$L_X$--$N_{200}$ relation as is appropriate to our catalog and stacking method.
This scatter comes about not only due to intrinsic $L_X$ variations between the
clusters (including different morphologies and merging clusters), but is also
due to point sources, cooling flows, and the projection of multiple clusters
along the line of sight.  For the $\sim 1000$ richest clusters ($N_{200} \geq
30$) used in this exercise, the chance of projection is very low.  We do not
have any way to remove point sources or bright cool cores from individual
clusters in an unbiased way.  Operationally, it is not relevant to our
measurement whether the scatter is due to intrinsic $L_X$ variations or due to
point source and cool-core contamination.  To constrain the observed scatter,
we need only assume that the underlying distribution is approximately
log-normal, which is consistent with our observations.

\citet{gbcz04} measure the observed scatter in the $L_X$--$\Bgc$ relation,
where $\Bgc$ is a richness measure from the amplitude of galaxy-cluster
correlation function~\citep{yl99}.  They use 290 optically-selected clusters,
of which 40 have significant detections in \emph{ROSAT}/PSPC observations.
Using a Bayesian maximum-likelihood fitting technique, they find a significant
correlation between $L_X$ (bolometric) and $\Bgc$, with a power-law slope of
$\sim 1.6$ and an intrinsic scatter of $\sim 50\%$.  It must be noted that they
introduce a prior weighting, $P_\mathrm{prior}(\siglnl) \propto 1/\siglnl$,
which gives more weight to models with lower scatter.  Again, it is difficult
to directly compare our measurement of the slope and scatter to the
\citet{gbcz04} measurements, due to different richness quantities and selection
functions.  

A large intrinsic scatter in the $L_X$--$N_{200}$ relation can account for
optically-selected clusters that appear ``underluminous'' in X-rays as compared
to their optical
richness~\citep[e.g.][]{bbbec94,bcecb97,dsmlp02,lvob03,bpggg04,gbcz04,pbbr07}.
In particular, \citet{pbbr07} present a set of ``Abell X-ray Underluminous''
(AXU) clusters with masses determined from velocity dispersion measurements,
that are significantly dimmer in X-rays than expected.  In
Figure~\ref{fig:linmix} the seven AXU clusters that match to maxBCG
($0.1<z<0.3$, $N_{200} > 30$) are denoted with open squares.  These are all
significantly dimmer than the mean $L_X$--$N_{200}$ relation, but are fully
consistent with the observed scatter.  Velocity dispersion measurements of
maxBCG clusters~(B07) and comparisons to simulations~\citep{rwkme07} have shown
that there is considerable mass-mixing in a given $N_{200}$ bin.  The
low-luminosity clusters could be from lower-mass halos that are picked up as
moderate richness in the cluster finder.  However, \citet{pbbr07} argue that
the AXU clusters have low $L_X$ relative to their mass.  This could be due to
biases in masses calculated from velocity dispersions with small
samples~\citep[e.g.][]{bmbdd06}, or to large intrinsic scatter in the
$L_X$--$M_{200}$ relation~\citep[e.g.][]{sebsn06}.  Each of these possibilities
(or a combination of the two) are consistent with the observations.  We further
explore the $L_X$--$M_{200}$ relation of maxBCG clusters in a companion
letter~\citep{ryk07}.

We have reason to believe that the scatter is not fixed as a function of
richness.  For example, B07 find a significantly larger scatter in the velocity
dispersion-richness ($\smedfull$--$\nmean$) relation at low richness compared
to high richness.  However, the quality of our data for this exercise is not
adequate to constrain the change in the scatter as a function of richness.
Therefore, for the remainder of the paper we adopt this nominal value of
$\siglnl = 0.86\pm0.03$ to correct the stacked (mean) $\lmean$ to obtain the
underlying median $\lmed$ values.  This essentially requires us to multiply all
$\lmean$ values by a factor of $0.69\pm0.02$.

\subsection{\label{sec:xrayselect}Comparison to X-Ray Selected Clusters}

We can compare the $\lmed$--$\nmean$ relation for optically-selected maxBCG
clusters to that from X-ray selected clusters from the literature.  For this
exercise we chose to compare to the NORAS~\citep{bvhmg00} and 400 square
degree~\citep[400d: ][]{bvheq06} catalogs, both constructed from ROSAT data,
and thus measured in the same energy range as used in this work.  The NORAS
catalog is an X-ray flux limited cluster catalog constructed from the RASS
photon maps in the northern sky.  Although it is known to be only $\sim50\%$
complete, it provides a large sample of bright X-ray clusters that overlap the
maxBCG survey region.  The 400d catalog is a serendipitous X-ray flux limited
cluster catalog constructed from pointed ROSAT PSPC observations. The 400d
catalog covers fields sampled from the whole sky, and thus only $\sim 50\%$ of
the catalog overlaps the maxBCG survey region.

Our intention in this exercise is to obtain a baseline comparison of the X-ray
and optical richness properties of X-ray selected clusters to
optically-selected clusters. A somewhat different comparison of NORAS and
maxBCG, focused on testing the completeness of maxBCG, was performed in
\citet{kmawe07b}.

To obtain a list of clean matches between NORAS or 400d clusters and maxBCG
clusters, we require that the X-ray position and BCG position be matched within
$250\,\kpc$, and the redshift difference to be less than 0.05. This ensures
both that the probability of a false match is $\ll 1$ and that we are not
selecting, e.g., merging clusters where the X-ray and optical catalogs might
utilize different deblending schemes.  There are 89 NORAS clusters that match
maxBCG, with a median redshift of $\zmed=0.18$, and 53 400d clusters that match
maxBCG, with a median redshift of $\zmed=0.16$.  We convert all X-ray
luminosities to 0.1-2.4~keV in the rest frame with our adopted cosmology.  Note
that the $L_X$ values obtained in NORAS and 400d have been corrected for
aperture effects.  However, with the adopted $\beta$ parameters ($\beta = 2/3$
for NORAS, and $\beta > 0.6$ for 400d) these corrections produce at most a
$\sim10\%$ offset in $L_X$ compared to the luminosities obtained in our scaled
$R_{200}$ apertures.

\begin{figure}
\begin{center}
\scalebox{0.85}{\rotatebox{270}{\plotone{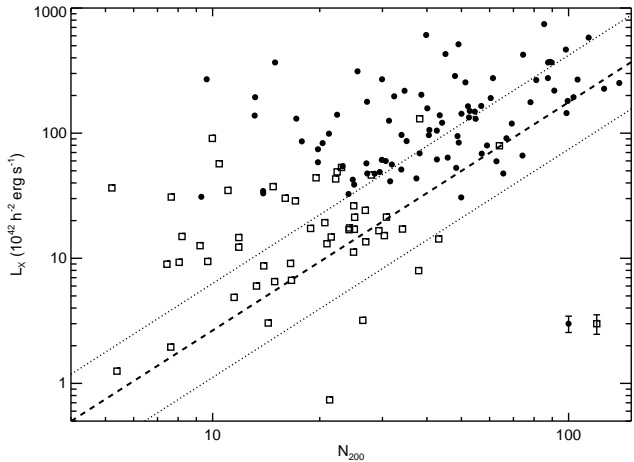}}}
\caption{\label{fig:catmatches}$L_X$ vs. $N_{200}$ for X-ray selected clusters
  that match the maxBCG catalog within $250\,\kpc$.  The 89 NORAS clusters
  (solid circles) are from a flux-limited survey using RASS, and the 53 400d
  clusters (empty squares) are from a deeper flux-limited survey using
  \emph{ROSAT}/PSPC pointed observations.  The dashed line shows the median
  $\lmed$--$\nmean$ relation, and the dotted lines show the $\pm1\siglnl$
  scatter constraints.  As seen in Figure~\ref{fig:linmix} there is significant
  scatter in the $L_X$--$N_{200}$ relation.  The X-ray selected clusters,
  especially from NORAS, tend to sample the brightest tail of the distribution
  at a given richness.}
\end{center}
\end{figure}

Figure~\ref{fig:catmatches} shows $L_X$ vs. $N_{200}$ for the X-ray selected
clusters that meet our matching criteria.  The solid circles represent NORAS
clusters, and the empty squares represent 400d clusters.  The luminosities of
the 400d clusters are typically less than that for NORAS due to the deeper flux
limit of the pointed ROSAT observations relative to the RASS survey.  The
dashed line shows the median $\lmed$--$\nmean$ relation, and the dotted lines
show the $\pm1\siglnl$ scatter constraints.  It is not surprising to note that
X-ray selection picks out primarily the X-ray brightest clusters at a given 
richness;
nearly all the NORAS clusters lie above the median maxBCG relation. The deeper
400d survey selects a sample more representative of the optically-selected
clusters, though it is still biased high. Comparison to Figure~\ref{fig:linmix}
is instructive.  At every richness, the typical X-ray emission from clusters is
significantly below what you might expect from X-ray selected catalogs. While
this is true for both catalogs, it is especially apparent for NORAS clusters.
Although the 400d survey has the sensitivity to detect the richest X-ray dim
clusters (as in the lower-right corner of Figure~\ref{fig:linmix}), these
clusters are quite rare.  The limited overlap between the 400d survey region
and the maxBCG survey region means that it is very unlikely that the 400d
survey would contain one of these specific maxBCG clusters.

\section{Biases in the $\lmed$--$\nmean$ Relation}
\label{sec:biases}

There are a variety of systematic effects which may bias the $\lmed$--$\nmean$ 
relation. In this section we address six: photometric redshift uncertainty, 
cluster centering errors, richness variation with redshift, BCG luminosity,  
point source contamination, and cool core clusters.

\subsection{Photometric Redshift Uncertainty}
\label{sec:photoz}

The maxBCG cluster catalog is based on photometric data, and thus we only have
photometric redshifts (``photo-$z$s'') for each of the clusters. While the
photo-$z$s are relatively precise ($\Delta_{z} \le 0.015$), we must investigate
how using photometric redshift estimates might bias our stacking results.

If a cluster is actually closer than its photometric redshift would suggest, it
will be over-weighted in the stacking analysis.  When it is more distant than
it seems, it will be underweighted.  Even if the photo-$z$ errors are perfectly
symmetric, the weighting factor, $w(z) = (d_L(z)/d_L(\zmed))^2$, is not.  As a
result, uncertainty in photo-$z$ can introduce a bias in the luminosities. 
This bias
can become significant if the photo-$z$ errors are large. The net impact of
this effect is to make the clusters appear more luminous than they truly are.

There is also a small volume effect caused by photo-$z$ errors. At a fixed
redshift $z$ there is a larger physical volume at a larger distance $z+\delta
z$ than at a smaller distance $z-\delta z$. Because of this, more clusters are
available at high redshift to scatter low than at low redshift to scatter
high. This is similar to Malmquist bias.  As above, this effect will cause us
to overestimate the cluster luminosities.

In addition, the photometric redshift is used to calculate the cluster
extraction radius.  When a cluster is closer than its photo-$z$ implies, then
the extraction radius used is too small, slightly reducing its implied
luminosity.  The opposite is true when the cluster is farther than its measured
photo-$z$.  As there is very little cluster flux at large radii, and as this
bias is roughly symmetric, we do not expect this effect to strongly bias the
stacked $L_X$.

We run a Monte Carlo simulation to estimate the combined photo-$z$ bias as
a function of richness.  Around $37\%$ of the BCGs in the maxBCG catalog used
in this analysis have spectroscopic redshifts from the SDSS DR5 spectroscopic
catalog~\citep{swlna02,eagsc01}.  With the spectroscopic subsample, we measure
the photo-$z$ offset ($\delta z = z_{ph}-z_{spec}$) distribution in each
richness bin and three redshift bins.  A Monte Carlo is performed to estimate
the overall bias in the measured luminosity by comparing the ``true''
luminosities convolved with the $\delta z$ distribution to the ``observed''
luminosities at the measured photometric redshifts.

 We find that the overall bias is small ($<6\%$ at 99\% confidence limit), 
with the
observed luminosities slightly overestimating the true luminosities.  This bias
factor does not vary with richness.  We neglect this correction factor due to
its small magnitude.

\subsection{\label{sec:centering}Centering Biases}

In our stacking exercise, we chose the BCG as the center of the cluster.
For many clusters with individual X-ray observations and a bright BCG, this is
coincident with the X-ray center.  However, there are several types of clusters
in which the BCG is not consistent with the center of the X-ray emission.
These include merging clusters with disturbed morphologies; clusters without a
dominant central galaxy for unambiguous detection by the maxBCG algorithm; and
a few ($\sim5$) previously identified strong cooling flow clusters. The central
galaxies in these clusters host bright AGN and exhibit strong star formation, 
moving their colors off the red sequence and excluding them from
the maxBCG cluster finding algorithm.

We do not expect centering biases to be a significant problem for our
calculation of the mean X-ray luminosity for the following reasons.  First, the
X-ray luminosity is proportional to the electron density $\rho^2$, and thus
most of the luminosity comes from the central region.  With our large apertures
($> 750\,\kpc$) as long as the core of a cluster is within the aperture, the
photon counting exercise will include most of the cluster photons.  Second, we
have reason to believe from simulations that most of the richer clusters are
well centered~(J07).  This is not to say that the cluster centering
will not have any effect: as shown in Section~\ref{sec:profiles}, the
$\beta$-model parameter is highly dependent on the centering distribution.

To confirm this, we perform some simple tests of de-centering the
clusters.  If we randomly offset all the cluster positions with a 2D Gaussian
with an rms of $300\,\kpc$~(comparable to the centering distribution
modeled in J07), the mean $\lmean$ values decrease by $< 10\%$, and less
than $1\sigma$.  This rms value was chosen because, as is seen in
Figure~\ref{fig:ptmatchhist}, the vast majority of the X-ray cluster matches
are within $250\,\kpc$.  Therefore, we do not consider cluster miscentering to
contribute a significant bias to our X-ray luminosity calculation.

\subsection{\label{sec:redshift}Richness Variation with Redshift}

We investigate whether the observed $\lmean$ at fixed richness changes with
redshift. Modest evolution in $\lmean$ at fixed \emph{mass} is expected. 
If the clusters are evolving in a self-similar manner, then we
expect the higher redshift clusters to be more luminous than the lower redshift
clusters, due to the fact that the Universe was more dense at higher redshift.
The expectation is that $L_X \propto \rho_{c}(z)^{7/6}$, where $\rho_{c}$ is
the critical density of the Universe at redshift $z$~\citep{k86}.  Therefore,
clusters at a given \emph{mass} at a redshift $z=0.3$ should be 
$\sim 30\%$ brighter than similar clusters at a redshift of $z=0.1$.

To study possible variation of $\lmean$ at fixed richness, 
we split the cluster sample into three redshift bins, containing the bottom
25\%, middle 50\%, and top 25\% of clusters in our sample.  The low redshift
bin ranges from $0.10 < z < 0.17$ with a median redshift of $\zmed=0.14$; the
middle redshift bin ranges from $0.17 < z < 0.26$ with a median redshift of
$\zmed=0.23$; and the high redshift bin ranges from $0.26 < z < 0.30$ with a median
of $\zmed=0.28$.  To obtain increased signal-to-noise in these bins, we 
combine richness bins $41 \leq N_{200} \leq 50$ and $51 \leq N_{200} \leq 70$;
$26 \leq N_{200} \leq 32$ and $33 \leq N_{200} \leq 40$; and $18 \leq N_{200}
\leq 20$ and $21 \leq N_{200} \leq 25$.  These wider bins are the same that
were used in the lensing analysis of S07.

\begin{figure}
\begin{center}
\scalebox{0.85}{\rotatebox{270}{\plotone{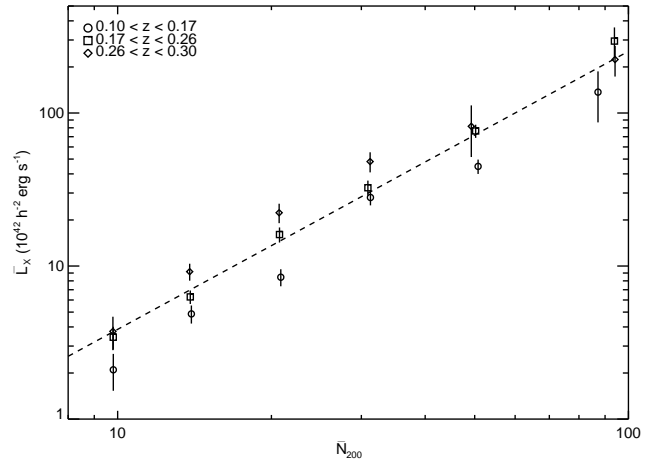}}}
\caption{\label{fig:splitzplot}Mean $\lmean$--$\nmean$ in three redshift bins,
  calculated within the scaled $R_{200}$ apertures.  The high redshift bin
  (diamonds) are significantly more luminous than the low richness bin
  (circles), indicating significant variation in the $\lmean$--$\nmean$
  relation with richness.  The dashed line is the stacked mean relation from
  Equation~\ref{eqn:lxn_r200}, which is consistent with the mean relation for
  the central redshift bin (squares).
}
\end{center}
\end{figure}

Figure~\ref{fig:splitzplot} shows the stacked $\lmean$ as a function of
$\nmean$ for the three different richness bins.  The dashed line shows the mean
relation (Equation~\ref{eqn:lxn_r200}) for all clusters from
Section~\ref{sec:meanrelation}.  The high redshift clusters (diamonds) are
significantly more luminous than the low redshift clusters (circles).  We
parameterize the variation with a factor of $(1+z)^\gamma$, and fit all the
redshift and richness bins simultaneously with a model of the form:
\begin{equation}
\label{eqn:zgamma}
\lmean(<R_{200}) = e^{\alpha} \left ( \frac{\nmean}{40} \right )^\beta \left (
\frac{1+z}{1+\zmed} \right )^\gamma\,10^{42}\,\lum,
\end{equation}
where $\zmed=0.23$, the median redshift of the cluster catalog.  This results
in a good fit ($\chi^2/\nu = 18.4/15$) with the best-fit parameters: $\alpha =
3.90 \pm 0.04$; $\beta = 1.85 \pm 0.05$; $\gamma = 6.0 \pm 0.8$.  Note that
$\alpha$ and $\beta$ are consistent with the mean relation in
Equation~\ref{eqn:lxn_r200}.  The redshift variation parameter, $\gamma$, is
quite large, and shows that the high redshift bin ($\zmed=0.28$) is almost
twice as bright as the low redshift bin ($\zmed=0.14$).  This is significantly
in excess of the self-similar prediction.  The challenge is to determine the
origin of this redshift dependent shift in $\lmean$ at fixed richness.

The photo-$z$ errors described in Section~\ref{sec:photoz} may account for some
of this effect. Although the absolute photo-$z$ errors are better
than $0.015$ at all redshifts, the \emph{relative} photo-$z$ errors are larger 
at low redshift than at high.  Therefore, the impact of photo-$z$ bias is
larger at low redshift than at high redshift.  We repeated the Monte Carlo
simulation described in Section~\ref{sec:photoz} to measure the photo-$z$ bias
from each redshift bin.  The additional redshift bias is $<10\%$ (99\%
confidence limit) in the low redshift bin, and $<2\%$ (99\% confidence limit)
in the high redshift bin.  The photo-$z$ bias therefore remains smaller than
the observed variation in $L_X$, although the effect is in the same sense as
the variation we observe, and is a contributing factor.

A more likely possibility is redshift dependent variation of $N_{200}$ at fixed
mass.  If our high redshift clusters have systematically smaller $N_{200}$ at
fixed mass, this would have the effect of shifting the high redshift points in
Figure~\ref{fig:splitzplot} to the left.  In order to constrain this bias, we
must first factor out the effect of the self-similar evolution of $\lmean$ at
fixed mass, using the expected redshift dependence $L_X \propto
\rho_{c}(z)^{7/6}$ as described above.  We then re-fit the data to
Equation~\ref{eqn:zgamma} to calculate the excess variation in the
$\lmean$--$\nmean$ relation that may be attributed to variation of $N_{200}$ at
fixed mass.  This results in a best-fit $\gamma=4.5\pm0.8$, implying a
fractional decrease in $N_{200}$ of $30\%$--$40\%$ from our lowest redshift bin
($\zmed = 0.14$) to our highest redshift bin ($\zmed = 0.28$).  This is
consistent with the redshift dependent variation in the velocity
dispersion--optical richness relation measured by B07.  However, we must note
that redshift dependence in the observed lensing shear, $\Delta\Sigma$, is
significantly smaller~(S07).  Unfortunately, none of these approaches is yet
able to confidently determine the nature of the observed variation. For
example, this effect could be caused by incorrect handling of the 0.4~$L_*$ and
color cuts that determine which galaxies are included in the richness
estimate. It might also be ``true'' evolution, such that clusters have fewer
red-sequence galaxies brighter than 0.4 $L_*$ at fixed mass at higher
redshifts. Further work, including improving the richness estimates, is in
progress to constrain the nature of this evolution.

\subsection{\label{sec:bcglum}BCG Luminosity}

We now investigate the effect BCG $i$-band luminosity ($\lbcg$, see
\citet{kmawe07b} for details) has on the mean X-ray luminosity $\lmean$.
Simulations and semi-analytical modeling predict that dark matter halos 
formed at early times have brighter BCGs and lower
richness than those which form late~\citep{zbbkw05,wzbka06,cgw07}.  Thus, 
we expect that the X-ray luminosity
should be correlated with $\lbcg$ at fixed richness.  Furthermore, optical and
X-ray observations of individual nearby clusters show that more luminous
BCGs are correlated with more massive halos with higher X-ray temperature and
luminosity~\citep[e.g.][]{lm04}.  This trend has already been seen in the mean 
velocity dispersion analysis of maxBCG clusters, as the clusters with more 
luminous BCGs had significantly larger velocity dispersions than clusters 
with similar richness~(B07).

\begin{figure}
\begin{center}
\scalebox{0.85}{\rotatebox{270}{\plotone{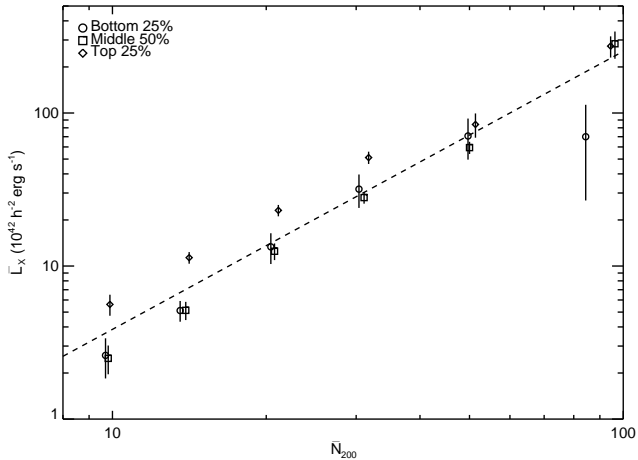}}}
\caption{\label{fig:splitbcgplot}Mean $\lmean$--$\nmean$ (within the scaled
  $R_{200}$ apertures) split according to BCG luminosity.  Each $N_{200}$ bin
  has been sorted by the optical luminosity of the BCG ($\lbcg$) of each
  cluster in the bin.  The stacking exercise has been performed on the top
  25\% (diamonds); middle 50\% (squares); and bottom 25\% (circles).  Note
  that $\lbcg$ is correlated with $N_{200}$, and thus the $\lbcg$ splits are
  different for each $N_{200}$ bin.  The dashed line shows the mean relation
  for all the clusters from Equation~\ref{eqn:lxn_r200}.  The clusters with the
  most luminous BCGs also tend to be the most luminous in X-rays.  This effect
  is strongest at low richness, where the BCG is a larger fraction of the total
  optical luminosity of the cluster.
}
\end{center}
\end{figure}

In an exercise similar to that performed in the previous section, we split
each wide richness bin into three bins of $\lbcg$.  As $\lbcg$, unlike
redshift, is correlated with $N_{200}$, we are unable to use the same $\lbcg$
split for each $N_{200}$ bin.  Therefore, we sort the clusters in each richness
bin by $\lbcg$, and split the sample into the top 25\%, middle 50\%, and bottom
25\%, and then restack.  The results are shown in
Figure~\ref{fig:splitbcgplot}.  It is readily apparent that the clusters with
the most luminous BCG in a given richness bin (diamonds) are significantly more
luminous in X-rays than other clusters in the bin.  This effect is most
dramatic at lower richnesses.  This is expected, as the BCG is more dominant in
low richness clusters than high richness clusters~\citep[e.g.][]{lm04}.  It is
also notable that there does not appear to be a significant difference between
the $\lmean$ for the low $\lbcg$ and middle $\lbcg$ bins.

We parameterize the effect of $\lbcg$ on $\lmean$ using a similar
procedure as performed for the redshift variation.  The richness and $\lbcg$
bins are fit simultaneously with a model of the form:
\begin{equation}
\lmean(<R_{200}) = e^{\alpha} \left ( \frac{\nmean}{40} \right )^\beta \left (
\frac{\lbcgmean}{10^{11}\,L_\sun} \right )^{\eta}\,10^{42}\,\lum.
\end{equation}
This results in an adequate fit ($\chi^2/\nu = 25.9/15$) with the best-fit
parameters: $\alpha = 3.86 \pm 0.04$; $\beta = 1.50 \pm 0.06$; $\eta = 0.82 \pm
0.10$.  Although the normalization is the same as in
Equation~\ref{eqn:lxn_r200}, the slope is significantly more shallow.  This is
caused by the correlation between $\lbcg$ and $N_{200}$.  After taking this
into account, the mean relation shown here is fully consistent with that
calculated without splitting the bins by $\lbcg$.  The reason the $\chi^2$ of
the fit is relatively high is because it does not appear that $\lmean$ is
smooth function of $\lbcgmean$: the effect of $\lbcg$ is not symmetric, and
only tends to boost the X-ray luminosity of clusters with relatively bright
BCGs.  

It is not entirely clear whether this effect is due to an imperfect cluster
finder, our crude richness definition, or due to real cluster physics.
Clusters with brighter, more dominant BCGs are more likely to be correctly
centered and thus might be more luminous in X-rays.  However, as we showed in
Section~\ref{sec:centering}, the effect of decentered clusters on $\lmean$ is
much smaller than the effect we see due to $\lbcg$.  Meanwhile, our richness
definition, $N_{200}$, is a count of red-sequence galaxies brighter than
$0.4\,L_*$. It is thus not surprising that a poor cluster with an exceptionally
bright BCG might be more massive and more X-ray luminous than a cluster at a
similar richness with a more typical BCG.  Finally, it is possible that these
clusters are associated with halos that formed early, allowing many of their
member galaxies to merge into a very large BCG~\citep{zbbkw05,wzbka06,cgw07}.  Deeper
targeted X-ray observations of specific clusters will be required to determine
if we can use $\lbcg$ and other optical properties to constrain the age of its
dark matter halo.

\subsection{\label{sec:ptsrc}Point Source Contamination}

When calculating the projected X-ray luminosity from the hot intracluster
medium, there is the possibility of contamination due to X-ray point sources.
Some point sources, such as cluster AGN, are associated with the cluster, but
in general their luminosities are not tightly correlated with that of the ICM.
Other point sources that are chance coincidences, such as foreground stars and
background quasars, might also boost the apparent luminosity.  The problem of
point source contamination is exacerbated by the broad PSF of the RASS survey.
Due to the large PSF, it is difficult to accurately excise point sources from
the higher redshift clusters, even when their positions are known.  For the
stacking exercise, foreground and background sources are not a significant
problem.  Assuming the positions of these point sources are uncorrelated with
the cluster positions, the contribution from these sources is subtracted with
the background estimation, as is demonstrated by our stack of random points in
Figure~\ref{fig:imageplots}.  At the same time, these point sources, both those
associated with the clusters and those that are chance projections, may
increase the observed scatter.  In general, we note that $L_X$ values
calculated from RASS are not corrected for point source
contamination~\citep[e.g.][]{bvhmg00, bsgcv04}.  We wish to perform a
first-order check on possible contamination from cluster AGN, the dominating
point sources that might bias the observed $\lmean$.

\citet{mmk07} have recently performed a detailed survey of the distribution of
AGN in galaxy clusters for eight moderate redshift clusters, comparable to the
redshift range of the maxBCG catalog.  They find that moderately bright AGN
($L_X > 10^{42}\,\mathrm{erg}\,\mathrm{s}^{-1}$) associated with bright cluster
members ($\gtrsim 0.5\,L_*$) make up around $\sim1\%$ of the total cluster
member population.  This implies that many of the richest maxBCG clusters
contain at least one moderately bright AGN.  These clusters have an extremely
bright ICM ($\lmean \sim 10^{44}\,\lum$), and thus the fraction of the
luminosity from cluster AGN will, on average, be very small.  If we extrapolate
the same AGN fraction to our poorest clusters ($N_{200} \sim 10$, perhaps one
in ten will host a moderately bright AGN.  Although these AGN might have a
luminosity comparable to the ICM for these $10\%$ of poor clusters, this small
fraction should not strongly bias the stacked average.

As a simple check for contamination in the X-ray signal from cluster AGN, we
match maxBCG cluster member galaxies to radio sources from the FIRST
survey~\citep{wbhg97} with a match radius of $3\arcsec$.  By matching random
galaxies from SDSS DR5 to the FIRST sources, we estimate the purity of the
matches to be $>99\%$.  The FIRST survey covers approximately the same
footprint as the SDSS, down to a typical flux limit of $1\,\mathrm{mJy}$ at 1.4
GHz.  Using 1.4 GHz radio detections as a proxy for AGN activity has the
advantage that the radio spectrum is often flat, and thus the 1.4 GHz flux is
not a strong function of redshift.  The disadvantages are that only 10\% of
X-ray bright quasars and AGN are radio loud, and that there is a large scatter
between X-ray and radio luminosities of these objects.  For example,
\citet{mmk07} noted that for the same small set of clusters none of the X-ray
selected AGN are radio loud~\citet{molkh03}.

Although we have an incomplete selection function, we can still constrain the
contamination from cluster AGN.  If the mean $\lmean$ is significantly boosted
by cluster AGN, then we would expect the clusters which match radio sources to
be relatively bright.  This effect would be greatest at low richness and low
$\lmean$ when the ICM is not as hot or bright.  For each richness bin, we
matched the maxBCG cluster members within $750\,\kpc$ of the BCG to sources in
the FIRST survey.  Table~\ref{tab:radiomatch} shows the number of clusters in
each bin that have members matched with the FIRST catalog.  More detailed work
is in progress in cross-correlating the maxBCG catalog with radio sources to
explore cluster AGN feedback and related issues (see also \citet{cdb07}).

\begin{deluxetable}{ccc}
\tablewidth{0pt}
\tablecaption{\label{tab:radiomatch}Radio Source maxBCG Member Matches}
\tablehead{
\colhead{Richness Range} & \colhead{Clusters w$/$} & \colhead{\% Clusters w$/$}\\
 & \colhead{Radio Sources} & \colhead{Radio Sources}
}
\startdata
$71 \leq N_{200} \leq 188$ & 45 & 82\%\\
$51 \leq N_{200} \leq 70$ & 91 & 65\%\\
$41 \leq N_{200} \leq 50$ & 125 & 62\%\\
$33 \leq N_{200} \leq 40$ & 183 & 54\%\\
$26 \leq N_{200} \leq 32$ & 284 & 45\%\\
$21 \leq N_{200} \leq 25$ & 420 & 37\%\\
$18 \leq N_{200} \leq 20$ & 332 & 30\%\\
$12 \leq N_{200} \leq 17$ & 1330 & 25\%\\
$9 \leq N_{200} \leq 11$ & 1341 & 18\%\\
\enddata
\end{deluxetable}

\begin{figure}
\begin{center}
\scalebox{0.85}{\rotatebox{270}{\plotone{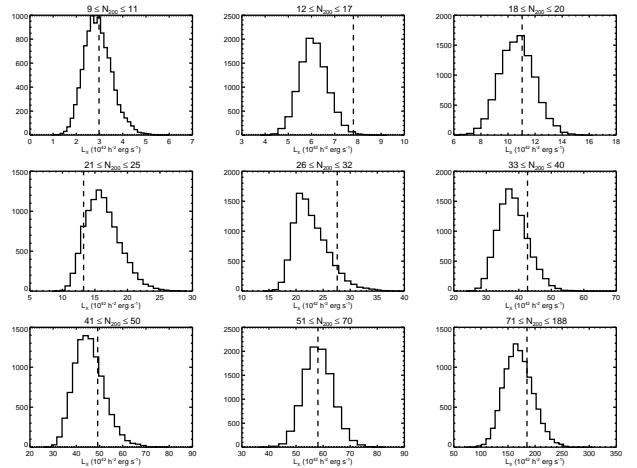}}}
\caption{\label{fig:radiomatch}Histograms of the distribution of $\lmean$
  values obtained from resampling the same number of clusters which match FIRST
  radio sources (as a proxy for AGN contamination) in each bin.  The vertical
  dashed lines indicate $\lmean$ for the clusters which match radio sources.
  The clusters with radio matches are not significantly more luminous than
  those drawn from the distribution, except for the $12 \leq N_{200} \leq 17$
  bin (discussed in the text).  This would indicate that, on average, AGN
  emission is not biasing our stacked $\lmean$ values.
}
\end{center}
\end{figure}

We wish to compare how the radio selected subset of clusters compares to a
randomly selected subset of clusters of similar richness.  In each richness
bin, we ran 10000 trials by sampling (with replacement) from the entire set of
clusters in the bin, sampling the same number of clusters that match radio
sources.  As a quick estimate of the mean luminosity, rather than performing
the full stacking analysis and spectral fits, we calculated the stacked and
weighted, background-subtracted hard channel (0.5-2.0 keV) counts in a fixed
$750\,\kpc$ aperture.  The resulting count rate was converted to $\lmean$ using
the method described in Section~\ref{sec:scatter}.  We can then compare the
mean $\lmean$ of the clusters which match radio sources to the typical values
of $\lmean$ that we would expect when drawing that same number of clusters from the
entire sample.  The results are shown in Figure~\ref{fig:radiomatch}.  In each
panel the histogram shows the distribution of $\lmean$ values obtained from the
resampling, and the vertical dashed line marks the mean $\lmean$ of the
radio-matched clusters. We do not see any significant bias as a function of
richness.  The only bin in which the radio matches appear to be outliers is the
$12\leq N_{200} \leq 17$ richness bin.  This is due to random chance, in that
the clusters in this richness bin with the highest $L_X$ also happen to match
radio sources; although there is no obvious indication in the RASS data that
there is point source contamination on these clusters, further follow-up with
\emph{Chandra} or \emph{XMM/Newton} would be required to clarify this.

\subsection{\label{sec:coolcore}Cool Core Clusters}

There is much evidence that cool core clusters increase the scatter in X-ray
cluster properties.  By excluding cluster cores in high resolution imaging,
many X-ray parameters become more tightly
correlated~\citep[e.g.][]{ombe06,crbiz07,m07}.  The broad ROSAT PSF, combined
with the fact that most of our clusters do not have individual X-ray
detections, makes it impossible to exclude cluster cores in our stacking
analysis.  However, a few of the brightest clusters ($>10^{44}\,\lum$) at 
moderate richness ($N_{200} \sim 45$) are well known cool-core clusters.  
We thus investigate what effect cool core clusters
have on our estimate of the median $\lmed$ and scatter $\siglnl$ in the
$\lmed$--$\nmean$ relation.

There does not exist an unbiased volume-limited catalog of cool core X-ray
clusters which can be fairly compared to the maxBCG catalog.  \citet{pfeaj98}
used ROSAT pointed observations of an X-ray flux limited catalog~\citep{esfa90}
to estimate the central cooling time of 55 nearby clusters.  Two of these
clusters are in the maxBCG catalog (A1689, A2244) and exhibit characteristics
of a cool core ($t_{cool} < 10\,\mathrm{Gyr}$).  \citet{bfsaj05} used
\emph{Chandra} observations of the higher redshift ($z>0.15$) BCS clusters to
systematically estimate the central cooling time of bright X-ray clusters.  We
identify 7 maxBCG clusters (A750, A963, A1835, RXJ 2129.6+0005, Z2701,
Z3146, Z7160) from this sample with cooling times $t_{cool} < 10\,\mathrm{Gyr}$
which we mark as cool core clusters.  Finally, we identify
ClG~J1504-0248~\citep{bbzsn05} as maxBCG cluster with a cool core.  It should
be noted that most of these clusters have a moderate richness ($30 < N_{200} <
50$) and are in the brightest tail of the $L_X$ distribution for their
richness.  Furthermore, the central galaxies of these clusters tend to show
strong $H\alpha$ emission in SDSS spectroscopy. In fact, for a few of these
clusters the true BCG is not in the list of cluster members because the strong
$H\alpha$ emission changes the apparent galaxy color, making it inconsistent
with the red-sequence used in the cluster finding algorithm.

After excluding these known cool core clusters (which might bias the
$\lmed$--$N_{200}$ relation) we recalculate the stacked mean relationship as
well as the median relationship with scatter.  When calculating the median
relationship with scatter, as in Section~\ref{sec:scatter}, the normalization
and slope shift by $\ll 1\sigma$, as these values are not strongly affected by
outliers.  However, the intrinsic observed scatter decreases to $\siglnl =
0.77\pm0.03$, which is a $\sim 2\sigma$ shift.  It is not surprising that
$\siglnl$ decreases: we are deliberately removing the brightest clusters.
Meanwhile, the stacked relation is also slightly affected by these bright
outliers, and after cutting the known cool core clusters the normalization of
the stacked relation decreases by $\sim 1\sigma$.  After following the
previously described prescription for converting the stacked mean relation to a
scatter-corrected median relation, we find that removing the known cool core
clusters decreases the normalization by $\lesssim 5\%$.  Therefore, it does not
appear that the known, very X-ray luminous, cool core clusters, are
significantly biasing our relation.  We reiterate that the intrinsic scatter we
measure is the \emph{observed} scatter that takes into account \emph{all} X-ray
observations of the maxBCG clusters, regardless of X-ray morphology of the
clusters.  Thus, it is likely that some portion of this observed scatter is
caused by unidentified cool core clusters in the sample.

\section{The Luminosity--Velocity Dispersion Relation
  ($\lmed$--$\smed$)}
\label{sec:lsig}

B07 have measured the median velocity dispersion ($\smed \equiv \smedfull$) as
a function of $\nmean$ for the maxBCG catalog.  The richness bins used to
measure the stacked velocity dispersions, as well as the scatter in the
$\sigma$--$N_{200}$ relation, are the same as those used in this work.
However, in order to measure velocity dispersions, only clusters that have the
BCG and at least one additional member galaxy in the SDSS DR5 spectroscopic
subsample~\citep{swlna02, eagsc01} were used.  The spectroscopic subsample 
preferentially selects bright/nearby galaxies, and thus the selection of 
clusters used in
B07 ($\zmed = 0.16$, with $z \geq 0.05$) is slightly different than that for
the full maxBCG catalog ($\zmed = 0.23$, with $z \geq 0.1$).  As discussed in
Section~\ref{sec:redshift}, the lower redshift clusters have a smaller $\lmean$
than the higher redshift clusters.  We therefore re-run our stacking
procedure on exactly those clusters used in B07, and use the known
spectroscopic redshifts of the BCGs rather than the photometric redshifts.
This ensures that, for this exercise at least, we are not affected by any
possible photometric redshift biases, which otherwise could be significant for
the nearest clusters ($z<0.1$) where the relative photo-$z$ error ($\Delta_z/z
\gtrsim 15\%$) is quite large.  Table~\ref{tab:lxsig} shows the richness bins
used, as well as the mean $\lmean$ (this work) and median $\smed$ (from B07)
for the nine richness bins.  The slope of the $\lmean$--$\nmean$ relation is
nearly identical for the spectroscopic subsample as reported in
Section~\ref{sec:meanrelation}, but the normalization is lower by $10\%$, due
to the different redshift selection.

\begin{deluxetable}{ccccc}
\tablewidth{0pt}
\tablecaption{\label{tab:lxsig}$\lmean$ and $\smed$ for Spectroscopic
  Subsample}
\tablehead{
\colhead{Richness Range} & \colhead{$\nmean$} & \colhead{$R_{200}$} &
\colhead{$\lmean$} & \colhead{$\smed$}\\
& & \colhead{($\Mpc$)} & \colhead{($10^{42}\,\lum$)} &
\colhead{$\mathrm{km}\,\mathrm{s}^{-1}$}
}
\startdata
$71 \leq N_{200} \leq 188$ & 83.9 & 1.69 & $131\pm28$ & $854\pm103$\\
$51 \leq N_{200} \leq 70$ & 58.4 & 1.45 & $86.6\pm11.4$ & $645\pm58$\\
$41 \leq N_{200} \leq 50$ & 44.7 & 1.32 & $57.2\pm6.9$ & $645\pm58$\\
$33 \leq N_{200} \leq 40$ & 35.9 & 1.22 & $40.3\pm4.9$ & $539\pm38$\\
$26 \leq N_{200} \leq 32$ & 28.6 & 1.10 & $24.5\pm2.5$ & $518\pm26$\\
$21 \leq N_{200} \leq 25$ & 22.7 & 1.02 & $15.2\pm2.8$ & $441\pm22$\\
$18 \leq N_{200} \leq 20$ & 19.0 & 0.94 & $9.05\pm1.39$ & $459\pm60$\\
$12 \leq N_{200} \leq 17$ & 14.0 & 0.82 & $6.95\pm0.84$ & $384\pm15$\\
$9 \leq N_{200} \leq 11$ & 9.9 & 0.73 & $2.94\pm0.62$ & $311\pm12$\\
\enddata
\end{deluxetable}

\begin{figure}
\begin{center}
\scalebox{0.85}{\rotatebox{270}{\plotone{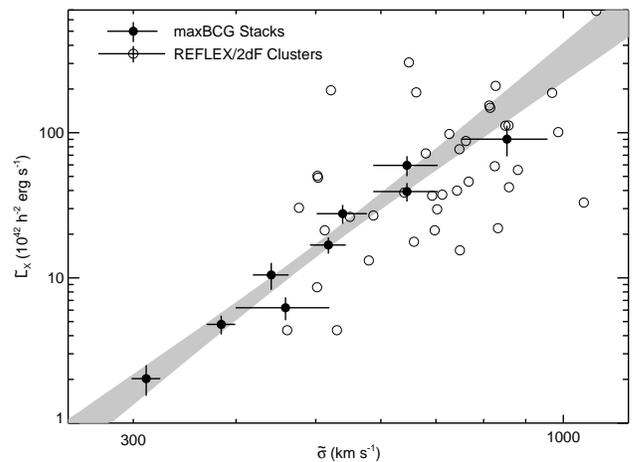}}}
\caption{\label{fig:lxsig}Median $\lmed$ vs. $\smed$ for the maxBCG clusters.
  $\lmed$ and $\smed$ for each stacked data point (solid circles) is measured
  in a given richness bin using the cluster selection from B07.  The gray band
  represents the best-fit relation ($\pm 1\sigma$), where $\lmed \propto
  \smed^{4.30\pm0.43}$, consistent with self-similar evolution.  The empty
  circles show $L_X$ and $\sigma$ values for individual X-ray selected REFLEX
  clusters as measured in the 2df survey~\citep{hcdbb05}, corrected to the
  cluster rest frame.  The fictitious data point in the legend in the upper
  left shows the typical errors on these values.  The individual X-ray selected
  clusters have slightly higher ($\sim 1\sigma$) velocity dispersions at a
  given X-ray luminosity, as compared to the maxBCG stacked measurements.  }
\end{center}
\end{figure}

We take the mean $\lmean$ values and corrected them for scatter as
described in Section~\ref{sec:scatter} to obtain $\lmed$.
Figure~\ref{fig:lxsig} shows the median $\lmed$--$\smed$ relation for maxBCG
clusters with richness $N_{200} \geq 9$.  Each data point (solid circles) is
obtained from the same richness bin and same cluster selection.  The dark gray
band shows the best-fit relation ($\pm 1\sigma$) from clusters in the
spectroscopic subsample of the maxBCG catalog, with the following functional
form:
\begin{equation}
\lmed (<R_{200}) = e^{2.73\pm0.11} \left (
\frac{\smed}{500\,\mathrm{km}\,\mathrm{s}^{-1}} \right )^{4.30\pm0.43}\,\lum
\end{equation}
The scaling relation predicted by self-similar cluster evolution is that $L_X
\propto T^2 \propto \sigma^4$~\citep[e.g.][]{k86,m98}, which is consistent with
our observations.  We note that if we could convert our 0.1-2.4~keV
luminosities to $\lxbolo$, this would steepen our relation, as the
clusters with higher velocity dispersions also have a higher temperature and
larger $\lxbolo/L_{X,0.1-2.4}$ ratio.  As mentioned previously, such
a conversion is not practical for the stacking exercise using ROSAT data.  The
slope of our 0.1-2.4 keV $\lmed$--$\smed$ relation is consistent with that
measured for REFLEX clusters~\citep{ogsbc04,hcdbb05}.

We compare this median relation to that obtained previously for individual
X-ray selected clusters.  \citet{hcdbb05} calculated the velocity dispersions
of REFLEX clusters by cross-correlating the REFLEX catalog with the Two-degree
Field Galaxy Redshift Survey~\citep[2dF;][]{cdmsn01}.  Figure~\ref{fig:lxsig}
shows 39 individual REFLEX clusters (empty circles) from Hilton, et al. (2007,
private communication) after converting the velocity dispersions to the cluster
rest frame.  In the upper-left corner the empty circle shows the typical errors
in 0.1-2.4 keV $L_X$ and $\sigma$ for the REFLEX/2dF clusters.  \citet{hcdbb05}
found the slope of the X-ray selected $L_X$--$\sigma$ relation to be
$4.0\pm0.6$, consistent with our observations.  However, the individual X-ray
selected clusters tend to have a slightly higher ($\sim 1\sigma$) velocity
dispersion at a given X-ray luminosity compared to the stacked maxBCG clusters.
This is possibly caused by the difficulty in measuring velocity dispersions
for individual clusters.  Contamination from non-virialized galaxies and
interlopers will more likely result in an overestimate of the velocity
dispersion rather than an underestimate.  

\section{Summary} \label{sec:summary}

By stacking observations from the \emph{ROSAT} All-Sky Survey, we measure the
mean X-ray luminosity as a function of richness for optically-selected clusters
from the maxBCG catalog.  With a large number of clusters in each bin,
the stacking exercise has the power to probe to much lower flux limits than are
possible for X-ray selected surveys. 
 Thus, a highly pure and complete volume-limited
catalog with clusters selected by their optical properties can be used to
measure the mean $\lmean$.

We find that $\lmean$ scales with optical richness, $N_{200}$, with a simple
power-law form over two orders of magnitude in $\lmean$.  Our results are
similar to those obtained in DKM07 by stacking RASS observations of clusters
identified in 2MASS.  However, we find that stacked temperature measurements
from \emph{ROSAT} are significantly biased for higher temperature clusters.
Therefore, we restrict our analysis to luminosities calculated in the
0.1-2.4~keV band rather than extrapolating to calculate bolometric
luminosities.  Furthermore, we have shown that the $\beta$-model fits to the
stacked radial profiles of moderate redshift clusters are dominated by a
combination of the broad \emph{ROSAT} PSF and the offset distribution between
maxBCG clusters and correlated X-ray sources.  On the other hand, calculating
X-ray luminosity in the 0.1-2.4 keV band, which relies primarily on counting
photons, is not strongly biased by spectral fits or profile fits.  As $L_X$
scales with the square of the density, as long as the core of a cluster is
within the stacking aperture, we can obtain a high reliability estimate of the
mean cluster luminosity.

By measuring the X-ray flux at the positions of individual maxBCG clusters, we
are able to constrain the observed scatter in the $L_X$--$N_{200}$ relation.
Assuming a log-normal distribution of $L_X$ as a function of $N_{200}$, we find
$\siglnl = 0.86\pm0.03$.  This scatter is quite large, and thus it is necessary
to correct for the scatter when calculating the median of the underlying
distribution $\lmed$.  The richness measure we have used in this work,
$N_{200}$, is simply the count of red-sequence galaxies within a scaled
$R_{200}$ aperture, but nevertheless is a good proxy for $L_X$ on average.  It
is likely that improved richness measurements can be made. From the evidence
presented here, it seems likely that these new richness estimators will 
include information about BCG luminosity, and will more carefully control 
variations in cluster membership with redshift.

The large scatter in the $L_X$--richness relation does have some significant
effects on cluster selection.  First, X-ray selection with a high flux limit,
as with the NORAS catalog, tends to pick out the clusters with the highest
$L_X$ at a given richness.  Second, the large scatter in the $L_X$--$N_{200}$ 
relation means that there are a significant number of optically selected 
clusters that are seemingly ``underluminous''. Optically selected clusters 
will often appear relatively dim compared to X-ray selected clusters at a
similar richness. 

The goal of cluster selection and identification is to find a low scatter proxy
for halo mass.  Both X-ray and optical selection techniques suffer from
different limitations.  X-ray flux limited surveys can only find the brightest
X-ray clusters, and are thus limited to a combination of the brightest and
nearest clusters.  Furthermore, there may be significant scatter between X-ray
luminosity and halo mass~\citep{sebsn06,nsre07}.  Optical surveys such as
maxBCG may be volume limited out to moderate redshift, but optical richness
also has a large scatter with respect to X-ray luminosity and mass~(B07).
The development of more precise mass proxies, including $T_X$, the Compton
Y parameter \citep{n06}, $Y_X$~\citep{kvn06}, and potentially more 
sophisticated optical richness estimates may allow selection of cluster 
catalogs more closely approximating the mass-limited catalog we aspire to.

By combining our present analysis with the velocity dispersion measurements of
B07, we have measured the median $\lmed$--$\smed$ relation for the maxBCG
clusters.  The relation has a slope of $4.30\pm0.43$, consistent with the
prediction of self-similar cluster evolution.  Previous measurements of the
$L_X$--$\sigma$ relation have been complicated by both selection effects and a
large scatter, as the derived slope depends strongly on the fitting technique
used.  By stacking clusters from a volume-limited catalog and correcting for
scatter, both $\lmed$ and $\smed$ are much better constrained, allowing a
robust calculation of the relation.  In our analysis, we do not see any
evidence of a break in the $\lmed$--$\smed$ relation at the poor cluster
(group) scale, as has been hinted at previously~\citep[e.g.][]{xw00}. This
might be due to the fact that the maxBCG cluster finder works with red-sequence
clusters, and does not find groups with a large blue galaxy fraction.

We can also compare $\lmean$ as determined from this method to weak lensing
masses determined from the same maxBCG cluster catalog~(S07, J07).
Determining the $L_X$--$M_{200}$ relation is not only important for
understanding cluster physics, but for calibrating the selection function of
X-ray clusters used as cosmological probes~\citep[e.g.][]{sebsn06}.  We have
made this measurement in a companion letter~\citep{ryk07}.  In
addition, the comparison of the X-ray luminosities, weak lensing masses, and
velocity dispersions can be used in conjunction with the number function of
maxBCG clusters to constrain cosmological parameters~\citep[e.g.][]{rwkme07}.
This work is ongoing.

Refined measurements of the scatter, point source contamination, and
confirmation of catalog purity will require deeper pointed X-ray observations.
We are currently investigating serendipitous maxBCG cluster observations by
deeper pointed \emph{ROSAT}/PSPC observations, as well as \emph{XMM/Newton} and
\emph{Chandra}~\citet[e.g.][]{pbgg05}.  Unfortunately, most of the
serendipitous cluster observations are of relatively poor clusters, which are
much more numerous.  A targeted campaign of a representative sample of maxBCG
clusters that have not previously been known to be X-ray bright will be
essential to both test our measurement of the scatter in $L_X$--$N_{200}$ and
to estimate the fraction of cluster flux that is contaminated by point sources.
This will be essential to enable future inexpensive deep optical cluster
surveys such as DES~\citep{des05} to use optical properties of clusters to
estimate which clusters would be most useful to follow-up with targeted X-ray
observations.

\acknowledgements 

E. Rykoff and T. McKay are pleased to acknowledge financial support from 
NSF AST-0206277 and AST-0407061, and the hospitality of the Michigan
Center for Theoretical Physics. MRB acknowledges the support of the Michigan
Space Grant Consortium.  AEE acknowledges support from NSF AST-0708150.  We
also wish to thank the anonymous referee for many helpful comments.

    Funding for the SDSS and SDSS-II has been provided by the Alfred P. 
Sloan Foundation, the Participating Institutions, the National Science 
Foundation, the U.S. Department of Energy, the National Aeronautics and 
Space Administration, the Japanese Monbukagakusho, the Max Planck Society, 
and the Higher Education Funding Council for England. The SDSS Web Site is 
http://www.sdss.org/.

    The SDSS is managed by the Astrophysical Research Consortium for the 
Participating Institutions. The Participating Institutions are the American 
Museum of Natural History, Astrophysical Institute Potsdam, University of 
Basel, University of Cambridge, Case Western Reserve University, University 
of Chicago, Drexel University, Fermilab, the Institute for Advanced Study, 
the Japan Participation Group, Johns Hopkins University, the Joint Institute 
for Nuclear Astrophysics, the Kavli Institute for Particle Astrophysics and 
Cosmology, the Korean Scientist Group, the Chinese Academy of 
Sciences (LAMOST), Los Alamos National Laboratory, the Max-Planck-Institute 
for Astronomy (MPIA), the Max-Planck-Institute for Astrophysics (MPA), New 
Mexico State University, Ohio State University, University of Pittsburgh, 
University of Portsmouth, Princeton University, the United States Naval 
Observatory, and the University of Washington.

\end{document}